\documentclass[10pt,journal]{IEEEtran}
\usepackage{cite}
\usepackage{graphicx}
\usepackage{amsmath}
\usepackage{times}
\usepackage{latexsym}
\usepackage{graphicx}
\usepackage{bm}
\usepackage{amssymb}
\usepackage[center]{caption2}
\usepackage{stfloats}
\usepackage{cases}
\usepackage{url}
\usepackage{array}
\usepackage{setspace}
\usepackage{fancyhdr}
\usepackage{color}
\usepackage{subfigure}
\usepackage{chemarrow}
\usepackage{extarrows}
\usepackage{algorithm}
\usepackage{algorithmic}

\newtheorem{theorem}{Theorem}
\newtheorem{lemma}{Lemma}
\newtheorem{corollary}{Corollary}
\newtheorem{proposition}{Proposition}

\def\proof{\noindent\hspace{2em}{\itshape Proof: }}
\def\endproof{\hspace*{\fill}~$\square$\par\endtrivlist\unskip}

\begin{document}
\title{Full-Duplex Massive MIMO Relaying Systems with Low-Resolution ADCs}
\author{Chuili Kong, \emph{Student Member}, \emph{IEEE}, Caijun Zhong, \emph{Senior Member}, \emph{IEEE}, Shi Jin, \emph{Member}, \emph{IEEE}, Sheng Yang, \emph{Member}, \emph{IEEE}, Hai Lin, \emph{Senior Member}, \emph{IEEE}, and Zhaoyang Zhang, \emph{Member}, \emph{IEEE}
\thanks{C. Kong, C. Zhong and Z. Zhang are with the Institute of Information and Communication Engineering, Zhejiang University, China. C. Zhong is also affiliated with the National Mobile Communications Research Laboratory, Southeast University, Nanjing, China (email: kcl\_dut@163.com; caijunzhong@zju.edu.cn; ning\_ming@zju.edu.cn).}
\thanks{S. Jin is with the National Mobile Communications Research Laboratory, Southeast University, Nanjing 210096, China (email:
jinshi@seu.edu.cn).}
\thanks{S. Yang is with the Laboratoire des Signaux et Syst\`emes, CentraleSup\'elec, Gif-sur-Yvette Cedex 91192, France (email: sheng.yang@supelec.fr).}
\thanks{H. Lin is with the Department of Electrical and Information Systems, Osaka Prefecture University, Osaka 599-8531, Japan (email: lin@eis.osakafu-u.ac.jp).}
}
\maketitle

\begin{abstract}

This paper considers a multipair amplify-and-forward massive MIMO relaying system with low-resolution ADCs at both the relay and destinations. The channel state information (CSI) at the relay is obtained via pilot training, which is then utilized to perform simple maximum-ratio combining/maximum-ratio transmission processing by the relay. Also, it is assumed that the destinations use statistical CSI to decode the transmitted signals. Exact and approximated closed-form expressions for the achievable sum rate are presented, which enable the efficient evaluation of the impact of key system parameters on the system performance. In addition, optimal relay power allocation scheme is studied, and power scaling law is characterized. It is found that, with only low-resolution ADCs at the relay, increasing the number of relay antennas is an effective method to compensate for the rate loss caused by coarse quantization. However, it becomes ineffective to handle the detrimental effect of low-resolution ADCs at the destination. Moreover, it is shown that deploying massive relay antenna arrays can still bring significant power savings, i.e., the transmit power of each source can be cut down proportional to $1/M$ to maintain a constant rate, where $M$ is the number of relay antennas.
\end{abstract}


\section{Introduction}\label{sec:introduction}
Full-duplex relaying has recently attracted considerable attention due to its substantial spectral efficiency gain over the conventional half-duplex relaying systems \cite{Q.Shi}. However, to realize the benefit, how to mitigate the loopback interference caused by the signal leakage from the relay output to its input, is one of the major issues to be tackled. Thus far, various techniques have been proposed to address this important issue, from separate transmit and receiver architecture to joint analog/digital filtering \cite{T.Riihonen}. In parallel, the massive multiple-input multiple-output (MIMO) technique is able to significantly boost the spectral efficiency and effectively suppress interference, hence has also received great interests recently \cite{T.L.Marzetta,H.Xie,H.Xie1,H.Xie2}. Therefore, it becomes a natural choice to combine these two promising technologies. The potential applications of such a massive antenna full-duplex relay include millimeter wave communications, device-to-device communications and machine-to-machine systems.

\subsection{Related Works and Motivation}
The performance of one-way and two-way multipair full-duplex relaying systems has been respectively studied in \cite{H.Q.Ngo2,X.Xia2} and \cite{Z.Zhang}, where it was demonstrated that deploying a large antenna array at the relay helps eliminate both the inter-pair and loop interference, thereby substantially boosting the achievable sum rate. However, such performance gain comes at the price of increased hardware cost and power consumption due to the extra required high-resolution analog-to-digital converters (ADCs), which may be undesirable in practical system deployment. The reasons are three-fold: 1) The power consumption of ADCs scales exponentially with the resolution and linearly with the sampling rate. For instance, a typical flash ADC with \emph{b}-bit and the sampling frequency $f_s$ operates $f_s 2^b$ conversion steps per second \cite{R.H.Walden}. 2) The fabrication cost of ADCs depends on the resolution. Thus, if high-resolution ADCs are adopted, the total financial cost will be a heavy burden for massive MIMO systems since each antenna requires a pair of ADCs to separately quantize the real and imaginary parts of signals. 3) The chip area of ADC increases exponentially with the resolution, which makes it difficult to put large number of collocated antenna arrays together. To resolve the above issues, a promising way is to use low-cost and power-efficient low-resolution or even one-bit ADCs unit to build radio frequency (RF) chains.

Low-resolution ADCs not only cause rate degradation, but also change some concluding remarks that have been made for unquantized systems. For example, the quality of the channel estimates depends on the set of orthogonal pilot sequences used, which is contrary to unquantized systems where any set of orthogonal pilot sequences gives the same result \cite{M.T.Ivrlac}. Moreover, compared to unquantized MIMO systems, the optimal length of training sequence is approximately 10 times more \cite{L.Fan2,Y.Li3}; while to achieve the same performance as that in a full channel state information (CSI) case, the length of pilot sequence increases to approximately 50 times the number of users \cite{C.Risi}, which is extremely long. Such a pilot overhead cannot be sustained and hence new channel estimation methods are proposed. For instance, \cite{C.-K.Wen} adopts a joint channel and data estimation approach to aid channel estimation and reduce pilot overhead. And other techniques such as expectation-maximization \cite{T.M.Lok,M.T.Ivrlac}, generalized approximate message passing \cite{J.Mo3}, and maximum likelihood algorithms \cite{J.Choi} have been proposed in the literature. Furthermore, the capacity achieving signals for single-input single-output system are no longer Gaussian distributions, and instead become discrete \cite{A.Mezghani2,J.Singh}. In addition, the detectors might need to be modified by taking the quantization effects into account.

Recently, there has been a surge of research interests to understand the impact of coarse quantization effects on massive MIMO systems. The works \cite{L.Fan,J.Zhang,L.Fan2,D.Qiao,D.Verenzuela,Y.Li,Y.Li1} have demonstrated that the massive antenna array has robustness against coarse quantization and capable of compensating for the performance loss caused by low-resolution ADCs. In particular, the work \cite{L.Fan2} studied the optimal training pilot length to maximize the spectral efficiency, while the work \cite{D.Verenzuela} revealed that the optimal number of quantization bits is 4 or 5 bits in terms of energy efficiency. For one-bit quantization systems, the power efficiency laws and energy-spectral efficiency tradeoff are characterized in \cite{Y.Li} and \cite{Y.Li1}, respectively. Moreover, various precoding methods such as spatio-temporal processing \cite{A.Gokceoglu}, minimum BER precoding \cite{H.Jedda}, and hybrid beamforming \cite{J.Mo} are studied. In addition, a mixed-ADC architecture is proposed to balance the spectral efficiency loss and power consumption \cite{N.Liang,C.Mollen,N.Liang2}. Despite the spectral efficiency enhancement, adopting mixed-ADCs increases the hardware complexity since an ADC switch is required. Furthermore, the optimal input symbol distribution and codebook for limited feedback have been designed \cite{J.Mo1,J.Mo2}. However, all the aforementioned works consider a single-hop system with low-resolution ADCs being implemented at either the BS or the destination users. Only very recently, \cite{J.Liu} introduces mixed-ADCs into the relay network, but it only considers the quantization at the BS. Therefore, the impact of low-resolution ADCs in a dual-hop system remains unknown.
\subsection{Our Work and Contributions}
Motivated by this, in this paper, we investigate the performance of full-duplex massive MIMO relaying systems with low-resolution ADCs at both the relay and the destinations. Specifically, we consider a multipair full-duplex relaying system using the amplify-and-forward (AF) protocol with simple maximum-ratio combining/maximum-ratio transmission (MRC/MRT) processing at the relay. First, CSI at the relay is acquired via pilot training, and the effect of low-resolution ADCs on the accuracy of CSI is characterized. Then, exact and approximated closed-form expressions for the sum rate are derived which enable efficient evaluation of the system's achievable sum rate. Moreover, based on the simple approximated sum rate expression, optimal relay power allocation strategy is characterized, and the power scaling law is studied. The findings of the paper suggest that, with only low-resolution ADCs at the relay, increasing the number of relay antennas is a promising method to compensate for the rate loss caused by the coarse quantization. However, compared to the infinite resolution ADC case, the required number of relay antennas doubles with one-bit ADCs. In addition, we show that the use of low-resolution ADCs at the destination is a major performance limiting factor, and it is preferable to deploy the low-resolution ADCs at the relay and use high-resolution ADCs at the destination. Finally, it is revealed that, even with low-resolution ADCs, deploying massive relay antenna arrays can still bring significant power savings, i.e., the transmit power of each source can be cut down proportional to $1/M$ to maintain a constant rate, where $M$ is the number of relay antennas.

\subsection{Organization and Notations}
The remainder of the paper is organized as follows: Section \ref{sec:system_model} introduces the multipair full-duplex relaying system model under consideration. Section \ref{sec:rate_analysis} presents an exact closed-form expression for the sum rate. Section \ref{sec:perfor_evaluation} provides an accurate approximation for the sum rate, and gives a detailed evaluation of the impact of low-resolution ADCs on the system performance. Numerical results are provided in Section \ref{sec:numerical_results}. Finally, Section \ref{sec:conclusion} summarizes the key findings.

{\it Notation}: We use bold upper case letters to denote matrices, bold lower case letters to denote vectors and lower case letters to denote scalars. Moreover, $(\cdot)^{H}$, $(\cdot)^{*}$, $(\cdot)^{T}$, and $(\cdot)^{-1}$ represent the conjugate transpose operator, the conjugate operator, the transpose operator, and the matrix inverse, respectively. Also, $|| \cdot ||$ is the Euclidian norm, $| \cdot |$ is the absolute value, and ${\left[ {\bf{A}} \right]_{mn}}$ gives the $(m,n)$-th entry of $\bf{A}$. In addition, ${\bf x} \thicksim {{\cal CN} ({\bf 0},{\bf \Sigma})}$ denotes a circularly symmetric complex Gaussian random vector ${\bf x}$ with zero mean and variance matrix ${\bf \Sigma}$, while ${{\bf{I}}_k}$ is the identity matrix of size $k$. Finally, the statistical expectation operator is represented by ${\tt E}\{\cdot\}$, the variance operator is ${\text{Var}} \left(\cdot\right)$, and the trace is denoted by ${\text{tr}}\left(\cdot\right)$.
\section{System Model}\label{sec:system_model}
\begin{figure}[!ht]
    \centering
    \includegraphics[scale=0.3]{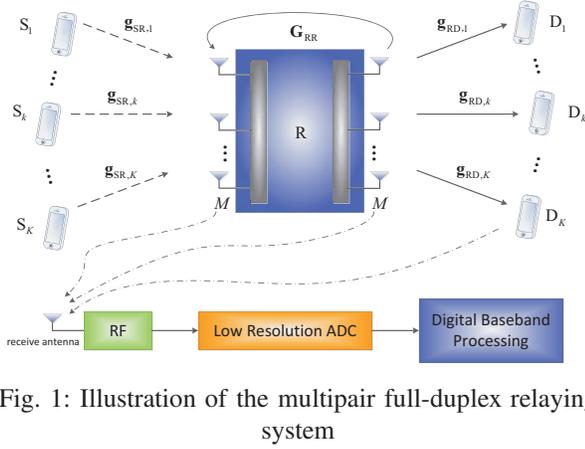}
    \caption{Illustration of the multipair full-duplex relaying system}\label{fig:system_model}
  \end{figure}

We consider a multipair AF relaying system shown in Fig. \ref{fig:system_model}, where $K$ single-antenna user pairs, denoted as ${\text S}_k$ and ${\text D}_{k}$, $k = 1,\ldots,K$, aim to exchange information with each other with the assistance of a shared multi-antenna full-duplex relay. To reduce the implementation cost, it is assumed that low-resolution ADCs are used at both the relay and destinations ${\text D}_{k}$ \footnote{Note that the considered model is very generic, and the single antenna destination node is not constrained to be the mobile phone, it could be certain low-cost device such as sensor, where using low-resolution ADC is desirable.}, $k = 1,\ldots,K$ \cite{J.Mo1,J.Mo2}. In addition, we assume that direct links between ${\text S}_{k}$ and ${\text D}_{k}$ do not exist due to large obstacles or severe shadowing \cite{M.Tao}.

During the $i$-th time slot, the sources ${\text S}_k$ $(k = 1,\ldots,K)$, transmit the signals $\sqrt{p_{\text S}}x_{{\text S},k}[i]$ satisfying ${\tt E}\left\{|x_{{\text S},k}[i]|^2 \right\} = 1$ to the relay, while the relay broadcasts the signal ${\bf x}_{\text R}[i]$ satisfying $ {\tt E}\left\{ {\bf x}_{\text R}[i] {\bf x}_{\text R}^H[i] \right\} = \frac{p_{\text R}}{M} {\bf I}_M$ simultaneously to all $K$ destinations ${\text D}_k$, $(k = 1,\ldots,K)$. Hence, the received signals at the relay and the $K$ destinations can be respectively expressed as
\begin{align}\label{eq:y_R}
  {\bf y}_{\text R}[i] &= \sqrt{p_{\text S}}{\bf G}_{\text{SR}} {\bf x}_{\text S}[i] + {\bf \bar{G}}_{\text{RR}} {\bf x}_{\text R}[i] + {\bf n}_{\text R}[i],\\
  {\bf y}_{\text D}[i] &= {\bf G}_{\text{RD}}^T {\bf x}_{\text R}[i] + {\bf n}_{\text D}[i],\label{eq:y_D}
\end{align}
where ${\bf x}_{\text S}[i] \triangleq \left[{ x}_{{\text S},1}[i], { x}_{{\text S},2}[i], \ldots, { x}_{{\text S},K}[i] \right]$. ${\bf G}_{\text{SR}} \in {\mathbb C}^{M \times K} $ and ${\bf G}_{\text{RD}} \in {\mathbb C}^{M \times K}$ denote the channels from the $K$ sources to the relay and from the relay to the $K$ destinations, respectively, which account for both small-scale and large-scale fading effects. More specifically, the \emph{k}-th columns of ${\bf G}_{\text{SR}} $ and ${\bf G}_{\text{RD}}$ are given by ${\bf g}_{{\text{SR}},k} \thicksim {\cal {CN}}\left({\bf 0}, \beta_{{\text{SR}},k}{\bf I}_M\right)$ and ${\bf g}_{{\text{RD}},k} \thicksim {\cal {CN}}\left({\bf 0}, \beta_{{\text{RD}},k}{\bf I}_M\right)$, where ${\beta_{{\text{SR}},k}}$ and ${\beta_{{\text{RD}},k}}$ model the large-scale path-loss effect, which are assumed to remain constant over many coherence intervals and known a priori. Also, ${\bf \bar{G}}_{\text{RR}} \in {\mathbb C}^{M \times M}$ represents the loop interference channel at the full-duplex relay. In addition, ${\bf n}_{\text R}[i]$ and ${\bf n}_{\text D}[i]$ denote the additive white Gaussian noise (AWGN) at the relay and the $K$ destinations, respectively. The elements of ${\bf n}_{\text R}[i]$ and ${\bf n}_{\text D}[i]$ are assumed to be independent and identically distributed (i.i.d.) ${\cal{CN}}\left(0,1\right)$.

\begin{figure}[!ht]
    \centering
    \includegraphics[scale=0.5]{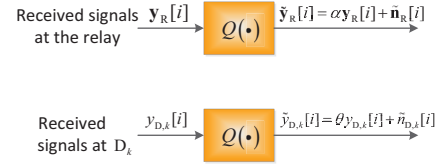}
    \caption{Outputs of receivers with AQNM.}\label{fig:AQNM_model}
  \end{figure}

To model the receivers with low-resolution ADCs, we focus on the non-uniform quantizer, and adopt the additive quantization noise model (AQNM) as in \cite{L.Fan,J.Mo,A.Mezghani,J.Zhang} for tractable analysis, shown in Fig. \ref{fig:AQNM_model}. As such, the outputs of the ADCs corresponding to input ${\bf y}_{\text R}[i]$ at the relay and ${ y}_{{\text D},k}[i]$ (${ y}_{{\text D},k}[i]$ is the \emph{k}-th element of ${\bf y}_{{\text D},k}[i]$) at ${\text D}_k$ are denoted as
\begin{align}\label{eq:tilde_y_R}
  {\tilde{\bf y}}_{\text R}[i] &= \alpha {\bf y}_{\text R}[i] + {\tilde{\bf n}}_{\text R}[i],\\
  {\tilde{ y}}_{{\text D},k}[i] &= \theta { y}_{{\text D},k}[i] + {\tilde{ n}}_{{\text D},k}[i],\label{eq:tilde_y_D}
\end{align}
respectively, where parameters $\alpha$ and $\theta$ are determined by the number of quantization bits $b$ of ADCs, which indicate the resolution of ADC. For instance, $\alpha = \theta = 1$ implies perfect ADCs. For $b \leq 5$, the typical values of $\alpha$ and $\theta$ are listed in Table \ref{table:tab_1}, while for $b > 5$, they can be approximated by $\alpha\left(\text{or}\  \theta \right) = 1 - \frac{\pi\sqrt{3}}{2} 2^{-2b}$ \cite{J.Max}. Also, ${\tilde{\bf n}}_{\text R}[i]$ and ${\tilde{ n}}_{{\text D},k}[i]$ represent the additive Gaussian quantization noise at the relay and ${\text D}_k$, respectively. For a fixed channel realization ${\bf G}_\text{SR}$, ${\bf G}_\text{RD}$, and ${\bf {\bar G}}_\text{RR}$, the covariance of ${\tilde{\bf n}}_{\text R}[i]$ and ${\tilde{ n}}_{{\text D},k}[i]$ are respectively given by
\begin{align}
  {\bf R}_{{\tilde{\bf n}}_{\text R}[i]} &= \alpha \left(1 - \alpha \right) {\text{diag}} \left( {\tt E} \left\{ {\bf y}_{\text R}[i] {\bf y}_{\text R}[i]^H \right\} \right),\\
   {R}_{{\tilde{\bf n}}_{{\text D},k}[i]} &= \theta \left(1 - \theta \right) {\tt E} \left\{ |{ y}_{{\text D},k}[i]|^2 \right\}.
\end{align}

\begin{table}[t]
\caption{$\alpha$ and $\theta$ for different ADC quantization bits $b$.}
\label{table:tab_1}
\begin{center}
\begin{tabular}{cccccc}
\hline
$b$ & 1 & 2 & 3 & 4 & 5\\ \hline
$\alpha(\text{or}\ \theta)$ & 0.6366 &0.8825&0.96546&0.990503&0.997501 \\ \hline
\end{tabular}
\end{center}
\end{table}

Note that this AQNM model is realistic enough since the quantization noise variance not only depends on the number of quantization bit but also scales up with the received power. Also, \cite{J.Mo} has proved that this AQNM model is accurate at the low signal-to-noise-ratio regime in which our considered system is most likely to operate.

\subsection{Channel Estimation}
 We utilize pilots to estimate the channel matrices ${\bf G}_\text{SR}$ and ${\bf G}_\text{RD}$, which is a technique widely used in the literature \cite{F.Gao1,F.Gao2}. Therefore, during each coherence interval of length $\tau_c$ (in symbols), the channel training occupies $2 \tau_p$ symbols (the minimum length of $\tau_{\text p}$ equals to the number of users $K$). First, the destinations remain silent and all sources transmit simultaneously their mutually orthogonal pilot sequences ${\bf \Phi}_{\text S} \in {\mathbb C}^{K \times \tau_{\text p}}$ to the relay. After that, all destinations transmit their mutually orthogonal pilot sequences ${\bf \Phi}_{\text D} \in {\mathbb C}^{K \times \tau_{\text p}}$ to the relay whilst the destinations remain silent\footnote{Note that the channel training scheme where all sources and destinations transmit their pilots simultaneously can obtain the same channel estimation accuracy as that in our proposed training approach, but with increased computational complexity.}. For analytical tractability, we adopt the same pilot sequences as in \cite{M.T.Ivrlac}. Thus, the received signals at the relay's receive and transmit antenna arrays are \cite{H.Q.Ngo2}
 \begin{align}\label{eq:Yrp}
   {\bf Y}_{\text{rp}} &= \sqrt{\tau_{\text p} p_{\text p}} {\bf G}_{\text{SR}} {\bf \Phi}_{\text S} + {\bf N}_{\text{rp}},\\
   {\bf Y}_{\text{tp}} &= \sqrt{\tau_{\text p} p_{\text p}} {\bf{ G}}_{\text{RD}} {\bf \Phi}_{\text D} + {\bf N}_{\text{tp}},
 \end{align}
respectively, where $p_{\text p}$ is the transmit power of each pilot symbol; ${\bf N}_{\text{rp}}$ and ${\bf N}_{\text{tp}}$ denote the noise at the receive and transmit antenna arrays of the relay, respectively, with i.i.d. ${\cal{CN}}\left(0,1\right)$ elements.

With low-resolution ADC receivers, the resulting quantized signals at the relay's receive and transmit antenna arrays respectively read as
\begin{align}\label{eq:Yrp:tilde}
{\tilde{\bf Y}}_{\text{rp}} &= \alpha {\bf Y}_{\text{rp}} + {{\bf N}}_{\text{rq}},\\
{\tilde{\bf Y}}_{\text{tp}} &= \alpha {\bf Y}_{\text{tp}} + {{\bf N}}_{\text{tq}},
\end{align}
where ${{\bf N}}_{\text{rq}}$ and ${{\bf N}}_{\text{tq}}$ represent the quantization noise, whose covariance matrices are respectively given by
\begin{align}
  {\bf R}_{{{\bf N}}_{\text{rq}}} = {\tt E} \left\{{{\bf N}}_{\text{rq}}{{\bf N}}_{\text{rq}}^H \right\} = \alpha \left(1 - \alpha\right) {\text{diag}} \left( {\tt E} \left\{ {\bf Y}_{\text{rp}} {\bf Y}_{\text{rp}}^H \right\} \right) ,\\
  {\bf R}_{{{\bf N}}_{\text{tq}}} = {\tt E} \left\{{{\bf N}}_{\text{tq}}{{\bf N}}_{\text{tq}}^H \right\} = \alpha \left(1 - \alpha\right) {\text{diag}} \left( {\tt E} \left\{ {\bf Y}_{\text{tp}} {\bf Y}_{\text{tp}}^H \right\} \right).
\end{align}

Assuming that the relay employs the minimum mean-square-error (MMSE) estimator, the channel matrices ${\bf { G}}_{\text{SR}}$ and ${\bf { G}}_{\text{RD}}$ can be decomposed as
  \begin{align}\label{eq:esti:channel:matrix:G_SR}
  {\bf G}_{\text{SR}} &= {\hat{\bf G}}_{\text{SR}} + {\bf E}_{\text{SR}},\\
  {\bf G}_{\text{RD}} &= {\hat{\bf G}}_{\text{RD}} + {\bf E}_{\text{RD}},\label{eq:esti:channel:matrix:G_RD}
  \end{align}
  where ${\bf E}_{\text{SR}}$ and ${\bf E}_{\text{RD}}$ are the estimation error matrices of ${\bf { G}}_{\text{SR}}$ and ${\bf { G}}_{\text{RD}}$. Due to the orthogonality property of MMSE estimators and the fact that ${\hat{\bf G}}_\text{SR}$, ${\bf E}_\text{SR}$, ${\hat{\bf G}}_\text{RD}$, and ${\bf E}_\text{RD}$ are complex Gaussian distributed, these matrices are independent of each other. By rewriting \eqref{eq:esti:channel:matrix:G_SR} and \eqref{eq:esti:channel:matrix:G_RD} in vector form, we have
  \begin{align}\label{eq:esti:channel:vector}
  {\bf g}_{{\text{SR},k}} &= {\hat{\bf g}}_{{\text{SR},k}} + {\bf e}_{{\text{SR},k}},\\
  {\bf g}_{{\text{RD},k}} &= {\hat{\bf g}}_{{\text{RD},k}} + {\bf e}_{{\text{RD},k}},
  \end{align}
  where ${\hat{\bf g}}_{{\text{SR},k}} $, ${\bf e}_{{\text{SR},k}} $, ${\hat{\bf g}}_{{\text{RD},k}}$, and ${\bf e}_{{\text{RD},k}}$ are the \emph{k}-th columns of ${\bf {\hat G}}_{\text{SR}}$, ${\bf E}_{\text{SR}}$, ${\bf{\hat G}}_{\text{RD}}$, and ${\bf E}_{\text{RD}}$, respectively, which are mutually independent.

  The distributions of these vectors are given in the following lemma that is essential to the ensuing analysis.
\begin{lemma}\label{theorm:channel_esti}
  The elements of ${\hat{\bf g}}_{{\text{SR},k}} $, ${\bf e}_{{\text{SR},k}} $, ${\hat{\bf g}}_{{\text{RD},k}}$, and ${\bf e}_{{\text{RD},k}}$ are independent Gaussian random variables with zero mean, variance $\sigma_{\text{SR},k}^2 = \frac{\alpha \tau_{\text p} p_{\text p} \beta_{\text{SR},k}^2}{1 + \tau_{\text p} p_{\text p} \beta_{\text{SR},k}}$, ${\tilde\sigma}_{\text{SR},k}^2 = \frac{ \beta_{\text{SR},k} + \left(1 - \alpha\right) \tau_{\text p} p_{\text p} \beta_{\text{SR},k}^2}{1 + \tau_{\text p} p_{\text p} \beta_{\text{SR},k}}$, $\sigma_{\text{RD},k}^2 = \frac{\alpha \tau_{\text p} p_{\text p} \beta_{\text{RD},k}^2}{1 + \tau_{\text p} p_{\text p} \beta_{\text{RD},k}}$, and ${\tilde\sigma}_{\text{RD},k}^2 = \frac{ \beta_{\text{RD},k} + \left(1 - \alpha\right) \tau_{\text p} p_{\text p} \beta_{\text{RD},k}^2}{1 + \tau_{\text p} p_{\text p} \beta_{\text{RD},k}}$, respectively.
\end{lemma}
\proof
See Appendix \ref{app:theorm:channel_esti}.
\endproof
\subsection{Data Transmission}
During the $i$-th time slot, the relay amplifies the previously received symbol and broadcasts it to the $K$ destinations. Thus, we have
\begin{align}\label{eq:x_R}
  {\bf x}_{\text R}[i] = \gamma {\bf F} {\tilde{\bf y}}_{\text R}[i - d],
\end{align}
where ${\bf F} \in {\mathbb C}^{M \times M}$ is the linear processing matrix to be specified shortly, $\gamma$ is an amplification constant factor which is chosen to satisfy the power constraint at the relay, and $d \geq 1$ denotes the processing delay.

\subsubsection{Loop Interference}
Since the relay operates in the full-duplex mode, it suffers from loop interference. However, since the relay is aware of its transmitted signal, some form of self-interference mitigation method can be applied, such that the remaining self-interference is sufficiently weak and can be treated as additional noise ${\bf {\hat x}}_{\text R}[i]$ \cite{Z.Zhang,L.J.Rofriguez} with the same power constraint as ${\bf { x}}_{\text R}[i]$, i.e., ${\tt E}\left\{ {\bf {\hat x}}_{\text R}[i] {\bf {\hat x}}_{\text R}^H[i] \right\} = \frac{p_{\text R}}{M} {\bf I}_M$. Therefore, \eqref{eq:y_R} can be re-expressed as
 \begin{align}\label{eq:y_R_new}
   {\bf y}_{\text R}[i] &= \sqrt{p_{\text S}}{\bf G}_{\text{SR}} {\bf x}_{\text S}[i] + {\bf G}_{\text{RR}} {\bf {\hat x}}_{\text R}[i] + {\bf n}_{\text R}[i],
 \end{align}
where ${\bf G}_\text{RR}$ models the residual loopback interference channel due to imperfect cancellation \cite{T.Riihonen2}. As in \cite{H.Q.Ngo2,Z.Zhang}, we assume that the entries of ${\bf G}_\text{RR}$ are i.i.d. ${\cal{CN}}\left(0,\sigma_\text{LI}^2\right)$ with $\sigma_\text{LI}^2$ denoting the level of loop interference.
\subsubsection{Linear Processing}
The relay treats the channel estimates as the true channels for subsequent linear processing. With MRC/MRT processing, $\bf F$ is given by
\begin{align}
  {\bf F} = {\bf {\hat G}}_\text{RD}^* {\bf {\hat G}}_\text{SR}^H.
\end{align}

Recall that $\gamma$ is chosen to meet the power constraint at the relay, after some simple algebraic manipulations, $\gamma$ can be obtained as \eqref{eq:gamma}, shown on the top of the next page.
\begin{figure*}
\begin{align}\label{eq:gamma}
  \gamma = \sqrt{\frac{p_{\text R}}{\alpha^2 \left(p_{\text S} {\tt E}\left\{||{\bf F} {\bf G}_\text{SR}||^2\right\} + \frac{p_\text R}{M} {\tt E} \left\{||{\bf F} {\bf G}_\text{RR}||^2 \right\} + {\tt E} \left\{||{\bf F}||^2 \right\}\right) + {\tt E} \left\{||{\bf F} {\bf{\tilde n}}_\text{R}||^2 \right\} }}.
\end{align}
\hrule
\end{figure*}

\section{System Design Guidelines}\label{sec:rate_analysis}
This section investigates the achievable rate with low-resolution ADCs both at the relay and destinations. In particular, an exact closed-form expression is derived for the system's achievable rate. Furthermore, we provide an answer to the important question of where to deploy the low-resolution ADCs to attain best performance.

Substituting \eqref{eq:y_R_new}, \eqref{eq:x_R}, \eqref{eq:y_D}, and \eqref{eq:tilde_y_R} into \eqref{eq:tilde_y_D}, we have
\begin{align}
 {\tilde y}_{{\text D},k}[i] &= \alpha \theta \gamma \sqrt{p_\text S} {\bf g}_{{\text{RD}},k}^T {\bf F} {\bf g}_{{\text{SR}},k} x_{{\text S},k}[i-d] \\ \notag
 &+ \alpha \theta \gamma \sqrt{p_\text S} \sum\limits_{j \neq k} {\bf g}_{{\text{RD}},k}^T {\bf F} {\bf g}_{{\text{SR}},j} x_{{\text S},j}[i-d]\\ \notag
 &+ \alpha \theta \gamma {\bf g}_{{\text{RD}},k}^T {\bf F} {\bf G}_{\text{RR}} {\bf{\hat x}}_{{\text R}}[i-d] \\ \notag
 &+ \alpha \theta \gamma {\bf g}_{{\text{RD}},k}^T {\bf F} {\bf n}_{{\text R}}[i-d] \\ \notag
 &+ \theta \gamma {\bf g}_{{\text{RD}},k}^T {\bf F} {\bf{\tilde n}}_{{\text R}}[i-d] + \theta n_{{\text D},k}[i] + {\tilde n}_{{\text D},k}[i]. \\ \notag
\end{align}

We consider the realistic case where the $K$ destinations do not have access to the instantaneous CSI, which is a typical assumption in the massive MIMO literature since the dissemination of instantaneous CSI is extremely costly for very large antenna array\footnote{Note that there are two typical ways to obtain the instantaneous CSI, i.e., downlink training and feedback. However, both methods incur huge overheads, hence is highly undesirable.}. Hence, ${\text D}_k$ uses only statistical CSI to decode the desired signal \cite{H.Q.Ngo2,J.Hoydis}. Thus, we have
\begin{align}\label{eq:y_Dk}
   {\tilde y}_{{\text D},k}[i] &= \underbrace{\alpha \theta \gamma \sqrt{p_\text S} {\tt E} \left\{{\bf g}_{{\text{RD}},k}^T {\bf F} {\bf g}_{{\text{SR}},k} \right\} x_{{\text S},k}[i-d]}_\text{desired signal} + \underbrace{n_{k}^\text{eff}[i]}_\text{effective noise},
\end{align}
where $n_{k}^\text{eff}[i]$ is the effective noise given by
\begin{align}
   &n_{k}^\text{eff}[i] = \\ \notag
& \underbrace{\alpha \theta \gamma \sqrt{p_\text S} \left({\bf g}_{{\text{RD}},k}^T {\bf F} {\bf g}_{{\text{SR}},k} - {\tt E} \left\{ {\bf g}_{{\text{RD}},k}^T {\bf F} {\bf g}_{{\text{SR}},k} \right\} \right) x_{{\text S},k}[i-d]}_\text{estimation error} \\ \notag
   &+ \underbrace{\alpha \theta \gamma \sqrt{p_\text S} \sum\limits_{j \neq k} {\bf g}_{{\text{RD}},k}^T {\bf F} {\bf g}_{{\text{SR}},j} x_{{\text S},j}[i-d]}_\text{interpair interference} \\ \notag
 &+ \underbrace{\alpha \theta \gamma {\bf g}_{{\text{RD}},k}^T {\bf F} {\bf G}_{\text{RR}} {\bf{\hat x}}_{{\text R}}[i-d]}_\text{loop interference} \\ \notag
 &+ \underbrace{\alpha \theta \gamma {\bf g}_{{\text{RD}},k}^T {\bf F} {\bf n}_{{\text R}}[i-d]}_\text{ noise at the relay}
 + \underbrace{\theta \gamma {\bf g}_{{\text{RD}},k}^T {\bf F} {\bf{\tilde n}}_{{\text R}}[i-d]}_\text{quantization noise at the relay}\\ \notag
  &+ \underbrace{\theta n_{{\text D},k}[i]}_\text{noise at ${\text D}_k$} + \underbrace{{\tilde n}_{{\text D},k}[i]}_\text{quantization noise at ${\text D}_k$}. \\ \notag
\end{align}

Noticing that the ``desired signal'' and the ``effective noise'' in \eqref{eq:y_Dk} are uncorrelated, and capitalizing on the fact that the worst-case uncorrelated additive noise is independent Gaussian noise, we obtain the following achievable rate of $k$-th user.

\begin{theorem}\label{theorm:se_exact}
  With low-resolution ADCs at the relay and ${\text D}_k$, the achievable rate of $k$-th user is given by
  \begin{align}\label{eq:R_k}
  R_{k} = \frac{\tau_{\text c} - 2\tau_{\text p}} {\tau_{\text c}} \log_2\left(1 + \text{SINR}_k \right),
\end{align}
where
\begin{align}
  \text{SINR}_k = \frac{A_k}{B_k + C_k + D_k + E_k + F_k + G_k + H_k},
\end{align}
\end{theorem}
with
  \begin{align}
  { A}_k &= p_{\text S} M^4 \sigma_{\text{SR},k}^4 \sigma_{\text{RD},k}^4,\label{eqn:ak}\\
  {B}_k &= p_{\text S} M^3 \sigma_{\text{SR},k}^2 \sigma_{\text{RD},k}^2 \left(\beta_{\text{SR},k} \sigma_{\text{RD},k}^2 + \beta_{\text{RD},k} \sigma_{\text{SR},k}^2 \right)   \\ \notag
  &+ p_{\text S} M^2 \beta_{\text{SR},k} \beta_{\text{RD},k} \sum\limits_{n=1}^K \sigma_{\text{SR},n}^2 \sigma_{\text{RD},n}^2,\\
  C_k &= p_{\text S} M^3 \sum\limits_{j \neq k} \left( \sigma_{\text{SR},k}^2 \sigma_{\text{RD},k}^4 \beta_{\text{SR},j} + \beta_{\text{RD},k} \sigma_{\text{SR},j}^4 \sigma_{\text{RD},j}^2  \right)  \\ \notag
  & + p_{\text S} M^2 \sum\limits_{j \neq k} \beta_{\text{SR},j} \beta_{\text{RD},k} \sum\limits_{n\neq k,j}^K \sigma_{\text{SR},n}^2 \sigma_{\text{RD},n}^2\\\notag
  &+ p_{\text S} M^2 \sum\limits_{j \neq k} \beta_{\text{SR},j} \beta_{\text{RD},k} \sigma_{\text{SR},k}^2 \sigma_{\text{RD},k}^2 \\ \notag
  &+ p_{\text S} M^2 \sum\limits_{j \neq k} \beta_{\text{SR},j} \beta_{\text{RD},k} \sigma_{\text{SR},j}^2 \sigma_{\text{RD},j}^2, \\ \notag
    D_k &= M^2 \sigma_\text{LI}^2 p_\text R \left( M \sigma_{\text{SR},k}^2 \sigma_{\text{RD},k}^4 + \beta_{\text{RD},k} \sum\limits_{n=1}^K \sigma_{\text{SR},n}^2 \sigma_{\text{RD},n}^2 \right),\\
      E_k &= M^2 \left( M \sigma_{\text{SR},k}^2 \sigma_{\text{RD},k}^4 + \beta_{\text{RD},k} \sum\limits_{n=1}^K \sigma_{\text{SR},n}^2 \sigma_{\text{RD},n}^2 \right),\\
  F_k &= \frac{1-\alpha}{\alpha} M^3  p_{\text S} \sigma_{\text{SR},k}^2 \sigma_{\text{RD},k}^4 \left( \sigma_{\text{SR},k}^2 + \sum\limits_{i =1}^K \beta_{\text{SR},i} \right)\\ \notag
   & + \frac{1-\alpha}{\alpha} M^3 \left(p_\text R \sigma_\text{LI}^2 + 1\right) \sigma_{\text{SR},k}^2 \sigma_{\text{RD},k}^4\\ \notag
   &+ \frac{1-\alpha}{\alpha} M^2 \left(p_\text R \sigma_\text{LI}^2 + 1\right) \beta_{\text{RD},k} \sum\limits_{n=1}^K\sigma_{\text{SR},n}^2 \sigma_{\text{RD},n}^2 \\ \notag
   &+ \frac{1-\alpha}{\alpha} M^2 p_\text S \beta_{\text{RD},k} \sum\limits_{n=1}^K \sigma_{\text{SR},n}^2 \sigma_{\text{RD},n}^2 \left(\sigma_{\text{SR},n}^2 + \sum\limits_{i=1}^K\beta_{\text{SR},i}\right),\\
   G_k &= \frac{1}{\alpha^2 \gamma^2},
  \end{align}
  \begin{align}
      H_k &= \frac{1-\theta}{\theta} p_{\text S} M^3\sigma_{\text{SR},k}^2 \sigma_{\text{RD},k}^4 \left(M\sigma_{\text{SR},k}^2 + \sum\limits_{i=1}^K \beta_{\text{SR},i}\right) \\ \notag
  &+ \frac{1-\theta}{\theta} p_{\text S} M^2\sigma_{\text{SR},k}^2 \left(M\sigma_{\text{SR},k}^2 + \sum\limits_{i=1}^K \beta_{\text{SR},i}\right)\beta_{\text{RD},k} \sum\limits_{i=1}^K \sigma_{\text{RD},i}^2 \\ \notag
  &+  \frac{1-\theta}{\alpha \theta} M^3 \left(p_\text R \sigma_\text{LI}^2 + 1\right) \sigma_{\text{SR},k}^2 \sigma_{\text{RD},k}^4\\ \notag
  &+ \frac{1-\theta}{\alpha \theta} M^2 \left(p_\text R \sigma_\text{LI}^2 + 1\right) \beta_{\text{RD},k} \sum\limits_{n=1}^K \sigma_{\text{SR},n}^2 \sigma_{\text{RD},n}^2 \\ \notag
    &+  \frac{\left(1 - \alpha\right)\left(1-\theta\right)}{\alpha \theta} M^3  p_{\text S} \sigma_{\text{SR},k}^2 \sigma_{\text{RD},k}^4 \left( \sigma_{\text{SR},k}^2 + \sum\limits_{i =1}^K \beta_{\text{SR},i} \right)\\ \notag
   &+  \frac{\left(1 - \alpha\right)\left(1-\theta\right)}{\alpha \theta}  p_\text S M^2 \beta_{\text{RD},k} \sum\limits_{n=1}^K \sigma_{\text{SR},n}^4 \sigma_{\text{RD},n}^2 \\ \notag
   &+ \frac{\left(1 - \alpha\right)\left(1-\theta\right)}{\alpha \theta}  p_\text S M^2 \beta_{\text{RD},k} \sum\limits_{n=1}^K \sigma_{\text{SR},n}^2 \sigma_{\text{RD},n}^2 \sum\limits_{i=1}^K\beta_{\text{SR},i} \\ \notag
   &+ \frac{1-\theta}{\alpha^2 \theta \gamma^2},\label{eqn:hk}
  \end{align}
  where $\gamma$ is given by \eqref{eq:gamma:theor} (shown on the top of the next page).
  \begin{figure*}\label{eq:gamma:theor}
  \begin{align}
    \gamma = \frac{1}{M} \sqrt{\frac{p_\text R}{p_\text S \sum\limits_{i =1}^K \sigma_{\text{SR},i}^4 \sigma_{\text{RD},i}^2 \left(M \alpha^2 + \alpha \left(1 - \alpha\right) \right) + \alpha \sum\limits_{i =1}^K \sigma_{\text{SR},i}^2 \sigma_{\text{RD},i}^2 \left(p_\text S \sum\limits_{j =1}^K \beta_{\text{SR},j} + p_\text R \sigma_\text{LI}^2 + 1\right)}}.
  \end{align}
  \hrule
  \end{figure*}

  \proof
  See Appendix \ref{app:theorm:se_exact} \endproof

Theorem \ref{theorm:se_exact} presents an exact closed-form expression for the achievable rate of ${\text D}_k$, which is valid for arbitrary configuration of relay antenna number and user pairs, which enables efficient evaluation of the achievable rate. In addition, it also reveals the impact of key system parameters on the achievable rate. For instance, It can be observed that ${R}_k$ is an increasing function with respect to $M$, suggesting the benefits of deploying large antenna array at the relay. Furthermore, ${R}_k$ decreases with $K$, which is also intuitive since a larger number of user pairs results in more severe inter-pair interference. Moreover, ${ R}_k$ reduces if $\alpha$ and/or $\theta$ become small, indicating that using low-resolution ADCs at the relay and/or the destinations always degrades the achievable rate.
\section{Sum Rate Approximation}\label{sec:perfor_evaluation}
In the previous section, an exact expression has been derived for the achievable rate of $k$-th user. However, the expression is rather involved, and is not amenable for further manipulations. Motivated by this, we now present a relatively simple large-scale approximation for sufficiently large $M$. Based on which, the optimal relay power allocation scheme is studied and the asymptotic power scaling law is characterized.

\begin{proposition}\label{theorm:se_approx}
  With low-resolution ADCs at the relay and ${\text D}_k$, for sufficiently large $M$, $R_k$ can be accurately approximated by ${\tilde R}_k$, which is given by
  \begin{align}\label{eq:R_k_theor}
    {\tilde R}_k = \frac{\tau_{\text c} - 2\tau_{\text p}} {\tau_{\text c}} \log_2\left(1 + \widetilde{\text{SINR}} \right),
  \end{align}
  where
  \begin{align}
    \widetilde{\text{SINR}} = \frac{{\tilde A}_k}{{\tilde B}_k + {\tilde C}_k + {\tilde D}_k + {\tilde E}_k + {\tilde F}_k + {\tilde G}_k + {\tilde H}_k}
  \end{align}
with
\begin{align}
{\tilde A}_k &=  p_{\text S} M \sigma_{\text{SR},k}^2 \sigma_{\text{RD},k}^2, \\
{\tilde B}_k &=  p_{\text S} \left(\beta_{\text{SR},k} \sigma_{\text{RD},k}^2 + \beta_{\text{RD},k} \sigma_{\text{SR},k}^2 \right), \\
{\tilde C}_k &=  p_{\text S} \sum\limits_{j \neq k} \left(\sigma_{\text{RD},k}^2 \beta_{\text{SR},j} + \frac{\beta_{\text{RD},k} \sigma_{\text{SR},j}^4 \sigma_{\text{RD},j}^2}{\sigma_{\text{SR},k}^2 \sigma_{\text{RD},k}^2} \right), \\
{\tilde D}_k &=  p_{\text R} \sigma_\text{LI}^2 \sigma_{\text{RD},k}^2, \\
{\tilde E}_k &=  \sigma_{\text{RD},k}^2, \\
{\tilde F}_k &= \frac{1 - \alpha}{\alpha} \sigma_{\text{RD},k}^2 \left(p_{\text S}\left( \sigma_{\text{SR},k}^2 + \sum\limits_{i =1}^K \beta_{\text{SR},i} \right) + p_\text R \sigma_\text{LI}^2 + 1 \right), \\
{\tilde G}_k &= \frac{p_{\text S}}{p_{\text R} \sigma_{\text{SR},k}^2 \sigma_{\text{RD},k}^2} \sum\limits_{i = 1}^K \sigma_{\text{SR},i}^4 \sigma_{\text{RD},i}^2,\\
{\tilde H}_k &= \frac{1 - \theta}{\theta} p_{\text S} \left( M \sigma_{\text{SR},k}^2 \sigma_{\text{RD},k}^2 + \frac{\sigma_{\text{SR},k}^2 \beta_{\text{RD},k}}{\sigma_{\text{RD},k}^2}\sum\limits_{i = 1}^K \sigma_{\text{RD},i}^2\right) \\ \notag
&+ \frac{1 - \theta}{\theta} p_{\text S} \sigma_{\text{RD},k}^2 \left(\sum\limits_{i = 1}^K \beta_{\text{SR},i} + \frac{p_\text R \sigma_\text{LI}^2 + 1}{\alpha p_{\text S}} \right)\\ \notag
&+ \frac{\left(1 - \alpha\right)\left(1 - \theta\right)}{\alpha \theta} \sigma_{\text{RD},k}^2  p_{\text S}\left( \sigma_{\text{SR},k}^2 + \sum\limits_{i =1}^K \beta_{\text{SR},i} \right) \\ \notag
&+ \frac{\left(1 - \theta\right) p_\text S \sum\limits_{i=1}^K\sigma_{\text{SR},i}^4 \sigma_{\text{RD},i}^2 }{\theta p_\text R \sigma_{\text{SR},k}^2 \sigma_{\text{RD},k}^2}. \\ \notag
\end{align}
\end{proposition}
\proof
By ignoring the insignificant terms in the large $M$ regime in Equations from (\ref{eqn:ak}) to (\ref{eqn:hk}), the desired result can be obtained after some simple algebraic manipulations.
\endproof

Despite being obtained under the assumption of large $M$, the above approximation turns out to be rather accurate even for moderate number of antennas, i.e., $M = 64$, as will be shown in Section \ref{sec:numerical_results}. In addition, we observe that the quantization noise at the relay (corresponding to the term ${\tilde F}_k$) is a function with respect to $\alpha$ whereas the quantization noise at the destination (corresponding to the term ${\tilde H}_k$) depends on both $\alpha$ and $\theta$. This is expected since the quantization noise scales with the power of input signals and the quantization level of ADCs. For the quantization noise at the relay, the input signals are only quantized once by the low-resolution ADCs at the relay; while for the quantization noise at the destinations, the input signals are double-quantized by the low-resolution ADCs at the relay and destinations. Finally, we see that the quantization noise at the destination is the most significant term which has the same order as the desired signal (corresponding to the term ${\tilde A}_k$). Unlike the quantization noise at the relay which can be mitigated by exploiting the large antenna array, the quantization noise at the destination can not be effectively suppressed, hence, is the major performance limiting factor as shown in the following corollary.

\begin{corollary}\label{coro:rate_limit}
  As $M \rightarrow \infty$, the rate of \emph{k}-th user with low-resolution ADCs at both the relay and destinations converges to
  \begin{align}\label{eq:R_k_theor:limit}
    {\tilde R}_k \rightarrow \frac{\tau_\text c - 2 \tau_\text p}{\tau_\text c} \log_2 \left( 1 + \frac{\theta}{1 - \theta}\right).
  \end{align}
\end{corollary}
\proof Starting from Proposition 1, by keeping only the most significant terms in (\ref{eq:R_k_theor}), the desired result can be obtained after some simple algebraic manipulations.
\endproof

Corollary \ref{coro:rate_limit} indicates that, in the asymptotic large antenna array regime, the rate of \emph{k}-th user converges to a finite limit. In addition, the limit is independent of the resolution level of relay ADCs, and is only determined by the resolution level of destination ADCs, which indicates that using large antenna array at the relay cannot compensate for the rate loss caused by low-resolution ADCs at the destinations.
\subsection{Power scaling law}
In this subsection, we investigate the potential for power saving in the data transmission phase due to the deployment of very large antenna array at the relay, when the channel estimation accuracy is fixed. Specifically, we assume that $p_\text p$ is fixed, while $p_\text S = E_\text S/M$ and $p_\text R = E_\text R/M$, where $E_\text S$ and $E_\text R$ are constants that do not scale with $M$. Then, as $M\rightarrow \infty$, the rate of \emph{k}-th user is provided by the following corollary.
\begin{corollary}\label{coro:scaling}
  For fixed $p_\text p$, $E_\text S$ and $E_\text R$, if $p_\text S = E_\text S/M$ and $p_\text R = E_\text R/M$, as $M\rightarrow \infty$, we have
  \begin{align}
  {\tilde R}_k \rightarrow \frac{\tau_\text c - 2\tau_\text p}{\tau_\text c} \log_2\left(1 + \frac{\theta}{1-\theta + \frac{1}{\alpha E_\text S \sigma_{\text{SR},k}^2} + \frac{\sum\limits_{i=1}^K \sigma_{\text{SR},i}^4 \sigma_{\text{RD},i}^2}{E_\text R \sigma_{\text{SR},k}^4 \sigma_{\text{RD},k}^4} } \right).
\end{align}
\end{corollary}

Corollary \ref{coro:scaling} reveals a rather remarkable result, that despite the low-resolution ADCs at both the relay and destinations, the transmit power of each source and the relay can still be cut down proportionally to $1/M$ while maintaining a desired constant rate, which is the same as the case with ideal infinite-resolution ADCs.
\subsection{Power allocation}
From Proposition \ref{theorm:se_approx}, it is not difficult to observe that ${\tilde R}_k$ improves when the transmit power of sources $p_\text S$ increases. As such, each user should transmit at the maximum power. However, due to the existence of loopback interference, the relationship between ${\tilde R}_k$ and $p_\text R$ is more complicated. As can be readily shown that ${\tilde R}_k \rightarrow 0$ when $p_\text R \rightarrow 0$ or $p_\text R \rightarrow \infty$. Hence, using the maximum relay power does not necessarily yield the best performance. Therefore, we now optimize the relay transmit power with the objective of maximizing the achievable sum rate of the system. Specifically, the optimization problem can be formulated as
\begin{align}
  {\cal P}_1: \mathop{\text{maximize}}\limits_{p_\text R} \quad &\sum\limits_{k=1}^K {\tilde R}_k \\
  {\text{subject to}} \quad  & p_{\text R} \geq 0.
\end{align}

Due to the involved sum rate expression, analytical characterization of the optimal $p_\text R^*$ is intractable. However, the optimal solution can be efficiently obtained by one-dimensional search such as bisection method. To gain further insights, we consider the special homogeneous case where all links have the same large-scale fading, e.g., $\beta_{\text{SR},k} = \beta_{\text{RD},k} = 1$. And we have the following important corollary:
\begin{corollary}\label{coro:opt:pr}
  In a homogeneous communication setting, i.e., $\beta_{\text{SR},k} = \beta_{\text{RD},k} = 1$, the optimal relay transmit power is given by
  \begin{align}
    p_{\text R}^* = \sqrt{\frac{\alpha p_\text S K}{\sigma_\text{LI}^2}}.
  \end{align}
\end{corollary}
\proof
When $\beta_{\text{SR},k} = \beta_{\text{RD},k} = 1$, the problem ${\cal P}_1$ is equivalent to minimizing the function $f\left(p_\text R\right) = {b+c p_\text R + \frac{d}{p_\text R}}$, where $b = \sigma^2 \left(2K p_\text S + \frac{1-\alpha}{\alpha\theta}p_\text S\left(\sigma^2 + K\right) + \frac{1 - \alpha \theta}{\alpha \theta} + \frac{1 - \theta}{\theta}p_\text S \left(2K + M\sigma^2\right) \right)$, $c = \frac{\sigma^2\sigma_\text{LI}^2}{\alpha \theta}$, and $d = \frac{p_\text S K \sigma^2}{\theta}$ with $\sigma^2 = \frac{\alpha \tau_\text p p_\text p}{1 + \tau_\text p p_\text p}$. Since $f''(p_\text R) = 2dx^{-3} > 0$, $f(p_\text R)$ is a convex function. Thus, the optimal transmit power of the relay $p_\text R^*$ is obtained by solving $f'(p_\text R) = 0$.
\endproof

Corollary \ref{coro:opt:pr} shows that the optimal transmit relay power is a function of the resolution level of ADCs at the relay, the transmit power of users, the number of user pairs, and the residual loop interference power, but is independent of the channel estimation accuracy, resolution level of ADCs at the destination, and the relay antenna number. First of all, we see that less power should be used when the residual loopback interference is high. This is reasonable since increasing the relay power would result in higher residual loopback interference. In contrast, if $K$ becomes large, we should increase $p_\text R$ to serve the additional users.
Now, we turn our attention to the impact of the resolution of ADCs at the relay. It is observed that the optimal relay power increases as the resolution of relay ADCs improves. This is also intuitive, since higher-resolution ADCs result in less quantization errors, hence, the benefit of using large transmit power becomes more significant. For instance, considering the special cases with perfect ADCs and one-bit ADCs, i.e., $\alpha = 1$ and $\alpha = 0.6366$, the difference between the optimal $p_\text R^*$ of the two cases can be computed as $\Delta p_\text R^* = \left(1 - \sqrt{0.0366} \right)\sqrt{\frac{ p_\text S K}{\sigma_\text{LI}^2}} = 0.2 \sqrt{\frac{ p_\text S K}{\sigma_\text{LI}^2}}$, which implies $20\%$ less power for the case with one-bit ADCs.

\subsection{Deploying low-resolution ADCs at the relay or the destinations?}
At this point, it is also worth noting that there exists tradeoff between $\alpha$ and $\theta$, i.e., the rate ${\tilde R}_k$ may remain unchanged by jointly adjusting $\alpha$ and $\theta$, indicating that it is possible to increase the ADC resolution at the relay to compensate for the performance loss due to low resolution ADCs at the destination or vice versa.

Now, let us consider two extreme cases: 1) $\alpha = \alpha_1 \neq 1, \theta = 1$, namely, low-resolution ADCs at the relay and infinite resolution ADCs at the destination. 2) $\alpha = 1, \theta = \theta_2 \neq 1$, namely, infinite resolution ADCs at the relay and low-resolution ADCs at the destination, where $\alpha_1$ and $\theta_2$ respectively denote the quantization level of the two cases.

If $\alpha = \alpha_1 \neq 1, \theta = 1$, ${\tilde R}_k$ reduces to
  \begin{align}\label{eq:R_k:reso_at_relay}
    {\tilde R}_k^\text R = \frac{\tau_{\text c} - 2\tau_{\text p}} {\tau_{\text c}} \log_2\left(1 + \frac{{\tilde A}_k}{{\tilde B}_k + {\tilde C}_k + {\tilde D}_k + {\tilde E}_k + {\tilde F}_k + {\tilde G}_k } \right),
  \end{align}
which can be written in the following form
\begin{align}
{\tilde R}_k^\text R = \frac{\tau_{\text c} - 2\tau_{\text p}} {\tau_{\text c}} \log_2\left(1 + aM \right),
\end{align}
 where $a$ is a constant determined by other system parameters such as $\alpha$ but independent of $M$. This result suggests that deploying large antenna array at the relay is an effective method to compensate for the rate degradation caused by low-resolution ADCs at the relay.

If $\alpha = 1, \theta = \theta_2 \neq 1$, ${\tilde R}_k$ reduces to
  \begin{align}
    {\tilde R}_k^\text D = \frac{\tau_{\text c} - 2\tau_{\text p}} {\tau_{\text c}} \log_2\left(1 + \frac{{\tilde A}_k}{{\tilde B}_k + {\tilde C}_k + {\tilde D}_k + {\tilde E}_k + {\tilde G}_k + {\bar H}_k } \right),
  \end{align}
  where
  \begin{align}
  {\bar H}_k &= \frac{1 - \theta}{\theta} p_{\text S} \left( M \sigma_{\text{SR},k}^2 \sigma_{\text{RD},k}^2 + \frac{\sigma_{\text{SR},k}^2 \beta_{\text{RD},k}}{\sigma_{\text{RD},k}^2}\sum\limits_{i = 1}^K \sigma_{\text{RD},i}^2 \right) \\ \notag
  & +\frac{1 - \theta}{\theta} p_{\text S} \sigma_{\text{RD},k}^2 \left(\sum\limits_{i = 1}^K \beta_{\text{SR},i} + \frac{p_\text R \sigma_\text{LI}^2 + 1}{p_{\text S}} \right). \\ \notag
  \end{align}

In contrast to the previous case, both the denominator and numerator scales with $M$, indicating that increasing the number of relay antennas is ineffective if the performance bottleneck is due to low-resolution ADCs at the destination. The above observations reveal the asymmetric impact of low-resolution ADCs at the relay and destination, and also shed key design insights on how to allocate the ADCs to achieve optimal performance. Specifically, we have the following important result:
\begin{corollary}\label{coro:ADCs_compare}
If $\alpha_1=\theta_2$, the achievable rate of case 1 is larger than that of case 2, i.e., ${\tilde R}_k^\text R \geq {\tilde R}_k^\text D$.
\end{corollary}
\proof
Noticing that ${\tilde F}_k < {\bar H}_k$ as $\alpha_1 = \theta_2 \neq 1$ and ${\tilde F}_k = {\bar H}_k = 0$ when $\alpha_1 = \theta_2 = 1$, the desired result can be easily obtained.
\endproof

Corollary \ref{coro:ADCs_compare} suggests that, in terms of maximizing the achievable rate, it is preferable to deploy the low-resolution ADCs at the relay and infinite resolution ADCs at the destination.
\subsection{Comparison between full-duplex and half-duplex modes}
We now compare the performance of the full- and half-duplex modes. For the half-duplex mode, the sources and the relay separately occupy $\frac{\tau_\text c - 2\tau_\text p}{2\tau_\text c}$ time to transmit signals. For fair comparison, the transmit powers of each source and the relay in half-duplex mode should be twice of the powers in full-duplex mode, to ensure that the total energy spent in a coherence interval for both modes are the same. Therefore, the achievable rate of $k$-th user in half-duplex mode is given by
\begin{align}
    {\tilde R}_k^\text H = \frac{\tau_{\text c} - 2\tau_{\text p}} {2\tau_{\text c}} \log_2\left(1 + \frac{2{\tilde A}_k}{2{\tilde B}_k + 2{\tilde C}_k + {\tilde E}_k + {\hat F}_k + {\tilde G}_k + {\hat H}_k} \right),
  \end{align}
  where
  \begin{align}
  {\hat F}_k &= \frac{1 - \alpha}{\alpha} \sigma_{\text{RD},k}^2 \left(2p_{\text S}\left( \sigma_{\text{SR},k}^2 + \sum\limits_{i =1}^K \beta_{\text{SR},i} \right) + 1 \right), \\
  {\hat H}_k &= \frac{1 - \theta}{\theta} 2p_{\text S} \left( M \sigma_{\text{SR},k}^2 \sigma_{\text{RD},k}^2 + \frac{\sigma_{\text{SR},k}^2 \beta_{\text{RD},k}}{\sigma_{\text{RD},k}^2}\sum\limits_{i = 1}^K \sigma_{\text{RD},i}^2 \right) \\ \notag
  &+ \frac{1 - \theta}{\theta} 2p_{\text S} \sigma_{\text{RD},k}^2 \left(\sum\limits_{i = 1}^K \beta_{\text{SR},i} + \frac{1}{2 \alpha p_{\text S}} \right) \\ \notag
  &+ \frac{\left(1 - \alpha\right)\left(1 - \theta\right)}{\alpha \theta} \sigma_{\text{RD},k}^22 p_{\text S}\left( \sigma_{\text{SR},k}^2 + \sum\limits_{i =1}^K \beta_{\text{SR},i} \right) \\ \notag
  &+ \frac{\left(1 - \theta\right) p_\text S \sum\limits_{i=1}^K\sigma_{\text{SR},i}^4 \sigma_{\text{RD},i}^2 }{\theta p_\text R \sigma_{\text{SR},k}^2 \sigma_{\text{RD},k}^2}. \\ \notag
\end{align}
It is difficult to tell which mode is better, since the achievable system performance depends on various parameters such as the transmit powers, channel gains, the number of relay antennas, and the loop interference level. In particular, the loop interference level and the number of relay antennas play critical roles. If all the other parameters are fixed and only the loop interference level changes, the full-duplex mode outperforms the half-duplex mode when $\sigma_\text{LI}^2 \leq \sigma_\text{LI,0}^2$, where $\sigma_\text{LI,0}^2$ is the root of $\sum_{k=1}^K {\tilde R}_k = \sum_{k=1}^K {\tilde R}_k^\text H$. Similarly, if only the number of relay antennas changes, the full-duplex mode outperforms the half-duplex mode when $M \geq M_0$, where $M_0$ is the root of $\sum_{k=1}^K {\tilde R}_k = \sum_{k=1}^K {\tilde R}_k^\text H$.

\section{Numerical Results}\label{sec:numerical_results}
In this section, we present numerical results to validate the previous theoretical analysis. For all illustrative examples, the length of the coherence interval is $\tau_\text c = 196$ (symbols), chosen according to the LTE standard. The length of pilot sequence is $\tau_\text p = K$. Also, we set the large-scale fading coefficient $\beta_{\text{SR},k} = \beta_{\text{RD},k} = 1$ for simplicity.
\subsection{Validation of analytical results}
\begin{figure}[!ht]
    \centering
    \includegraphics[scale=0.5]{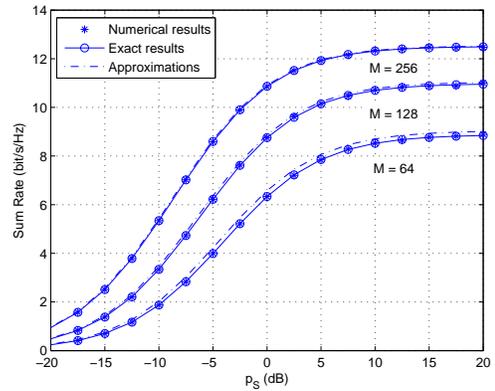}
    \caption{Sum rate versus $p_\text S$ for $K = 5$, $\alpha = \theta = 0.8825$ (two-bit ADC), $p_\text p = 10$ dB, $p_\text R = 10$ dB, and $\sigma_\text{LI}^2 = 0$ dB.}\label{fig:tightness_rate_ps}
  \end{figure}

Fig. \ref{fig:tightness_rate_ps} illustrates the impact of relay antenna number on the sum rate of the $K$ destinations. Note that the curves associated with ``Numerical results'' are generated by Monte-Carlo simulations according to \eqref{eq:R_k} by averaging over $10^4$ independent channel realizations, the ``Exact results'' curves are plotted according to Theorem \ref{theorm:se_exact}, and the ``Approximations'' curves are obtained based on Proposition \ref{theorm:se_approx}. As can be readily observed, the ``Exact results'' and ``Numerical results'' curves overlap, which validates our exact analysis. Also, the gap between ``Approximations'' and ``Numerical results'' curves is sufficiently small, especially when the number of relay antenna is large. In addition, we observe the sum rate saturates in the high signal-to-noise ratio (SNR) regime. Intuitively, the system becomes interference-limited at high SNR.

\subsection{Where to deploy the low-resolution ADCs?}
\begin{figure}[!ht]
    \centering
    \includegraphics[scale=0.5]{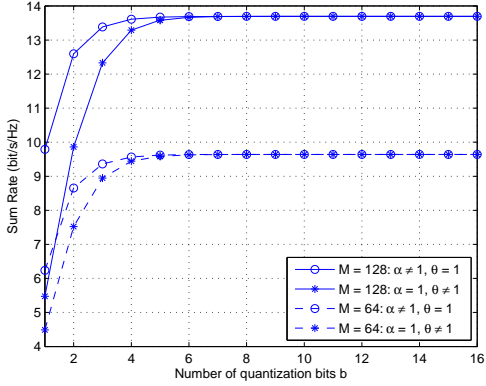}
    \caption{Sum rate versus $b$ with full-resolution ADCs at the relay or the destinations ($\alpha = 1 \ \text{or}\ \theta = 1$) for $K = 5$, $p_\text S = 0$ dB, $p_\text p = 0$ dB, $p_\text R = 0$ dB, and $\sigma_\text{LI}^2 = -10$ dB.}\label{fig:rate_quantization_bit_compare}
  \end{figure}

Fig. \ref{fig:rate_quantization_bit_compare} demonstrates the asymmetric effect of low-resolution ADCs at the relay and destinations on the sum rate of the $K$ destinations. Recall that $\alpha \neq 1,\theta = 1$ represents employing low-resolution ADCs at the relay only, while $\alpha = 1,\theta \neq 1$ refers to using low-resolution ADCs at the destinations only. We observe a substantial performance gap between the case with $\alpha \neq 1,\theta = 1$ and the case with $\alpha = 1,\theta \neq 1$ when the number of quantization bits $b$ is small, as predicted by Corollary \ref{coro:ADCs_compare}. However, when the number of quantization bits $b$ increases, the two curves converge to the same rate. This is because both $\alpha$ and $\theta$ approach to 1 as $b$ is large enough ($b \geq 6$ bits in this example), as such the system behaves as that with perfect ADCs.

\subsection{Can we use more antennas to compensate for the coarse quantization?}
\begin{figure}[!ht]
    \centering
    \includegraphics[scale=0.5]{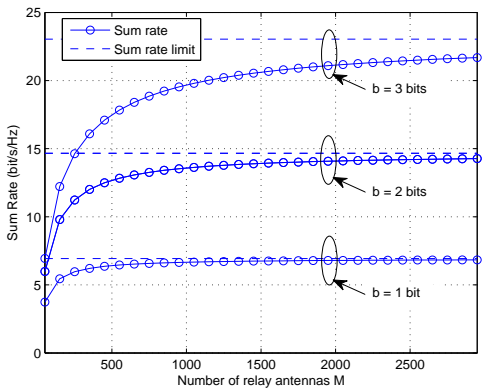}
    \caption{Sum rate versus $M$ for $K = 5$, $\alpha = 0.8825$, $p_\text S = 0$ dB, $p_\text p = 0$ dB, $p_\text R = 0$ dB, and $\sigma_\text{LI}^2 = -10$ dB.}\label{fig:rate_antenna_limits}
  \end{figure}

\begin{figure}[!ht]
    \centering
    \includegraphics[scale=0.5]{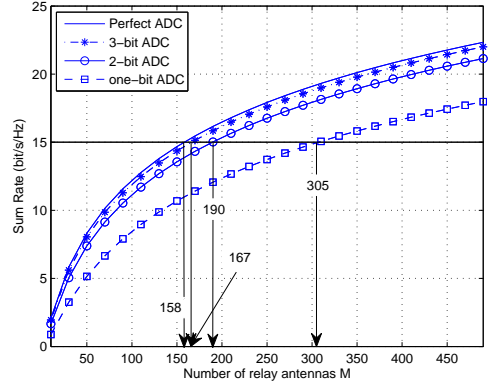}
    \caption{Sum rate versus $M$ for $K = 5$, $\theta = 1$, $p_\text S = 0$ dB, $p_\text p = 0$ dB, $p_\text R = 0$ dB, and $\sigma_\text{LI}^2 = -10$ dB.}\label{fig:rate_antenna_low_resol_relay}
  \end{figure}

Fig. \ref{fig:rate_antenna_limits} plots the sum rate of the $K$ destinations versus the number of relay antennas for different number of quantization bits at the destinations. Note that the ``Sum rate'' and ``Sum rate limit'' curves are generated by \eqref{eq:R_k} and \eqref{eq:R_k_theor:limit}, respectively. As expected, higher number of quantization bits of ADCs at the destinations results in better sum rate. Also, the sum rate is an increasing function with respect to the number of relay antennas, and converges to a finite limit determined by $b$ as $M$ becomes large as predicated by Corollary \ref{coro:rate_limit}, which indicates that using more relay antennas is not an effective approach to compensate for the rate loss due to low resolution ADCs at the destinations. However, the claim would be quite different in the case of low-resolution ADCs at the relay only. As illustrated in Fig. \ref{fig:rate_antenna_low_resol_relay}, increasing the number of relay antennas can effectively compensate for the rate degradation due to coarse quantization. Nevertheless, the required number of antennas is closely related to the resolution level of ADCs. For instance, compared to the perfect ADCs case, the one-bit system requires approximately twice ($305/158 = 1.93$) antennas to achieve a sum rate of $15$ bit/s/Hz (marked by a solid black line), while the 3-bit ADCs system merely needs an additional $9/158 = 5.7\%$ more antennas.

\subsection{How does coarse quantization affect the relay transmit power?}
\begin{figure}[!ht]
    \centering
    \includegraphics[scale=0.5]{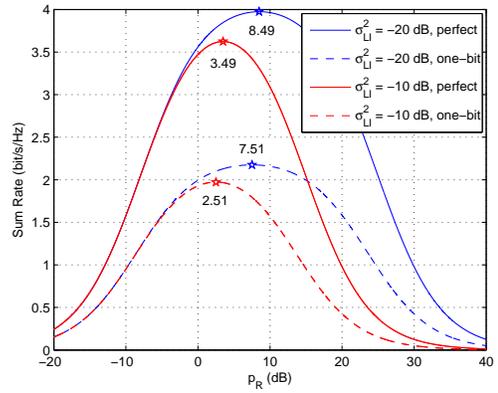}
    \caption{Sum rate versus $p_\text R$ for $K = 5$, $M = 64$, $\theta = 0.8825$, $p_\text S = -10$ dB, and $p_\text p = -10$ dB.}\label{fig:rate_pr}
  \end{figure}

Fig. \ref{fig:rate_pr} illustrates the impact of relay transmit power on the achievable sum rate of the $K$ destinations with different loopback interference level. As can be readily observed, there exists an optimal relay transmit power maximizing the sum rate. Also, with one-bit ADCs at the relay, the optimal relay transmit power decreases if the loop interference level increases ($p_\text R^* = 7.51$ dB for $\sigma_\text{LI}^2 = -20$ dB, and $p_\text R^* = 2.51$ dB for $\sigma_\text{LI}^2 = -10$ dB), which aligns with Corollary \ref{coro:opt:pr}. In addition, we can see that, regardless of the loop interference level, the relay with one-bit ADCs should transmit approximately $20\%$ less power than the perfect ADC case.

\subsection{Power scaling law}
\begin{figure}[!ht]
    \centering
    \includegraphics[scale=0.5]{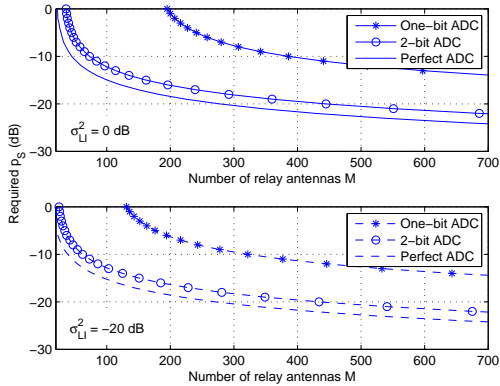}
    \caption{Required $p_\text S$ versus the number of relay antennas for $K = 5$, $p_\text p = 0$ dB, and $p_\text R = K p_\text S$.}\label{fig:ps_num_antenna}
  \end{figure}

Fig. \ref{fig:ps_num_antenna} plots the required transmit power of each source to maintain a given sum rate of the $K$ destinations 5 bits/s/Hz. Note that the curves associated with ``One-bit ADC'', ``2-bit ADC'' and ``Perfect ADC'' are obtained by setting $\alpha = \theta = 0.6366$, $\alpha = \theta = 0.8825$, and $\alpha = \theta = 1$, respectively. We can see that, when the number of relay antennas increases, the required transmit powers are significantly reduced. Also, lower-resolution ADC costs more power to achieve the target sum rate. For instance, compared to the perfect ADC case, the one-bit and two-bit ADCs require approximately 10 dB and 2.5 dB more power, respectively. In addition, when $\sigma_\text{LI}^2$ becomes large, we need more transmit power. For instance, with one-bit ADC and $M = 200$, the required transmit power of each source increases from $-6.25$ dB for $\sigma_\text{LI}^2 = -20$ dB to $-1.25$ dB for $\sigma_\text{LI}^2 = 0$ dB.

\subsection{Full-duplex vs. half-duplex modes}
\begin{figure}[!ht]
    \centering
    \includegraphics[scale=0.5]{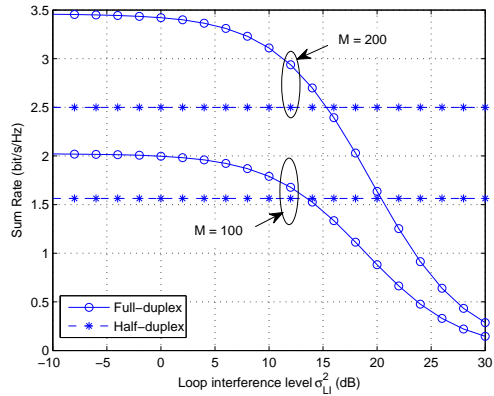}
    \caption{Sum rate versus $\sigma_\text{LI}^2$ for $K = 5$, $\alpha = \theta = 0.8825$, $p_\text p = -10$ dB, $p_\text S = -10$ dB, and $p_\text R = -10$ dB.}\label{fig:FD_vs_HD_loop_inter}
  \end{figure}

\begin{figure}[!ht]
    \centering
    \includegraphics[scale=0.5]{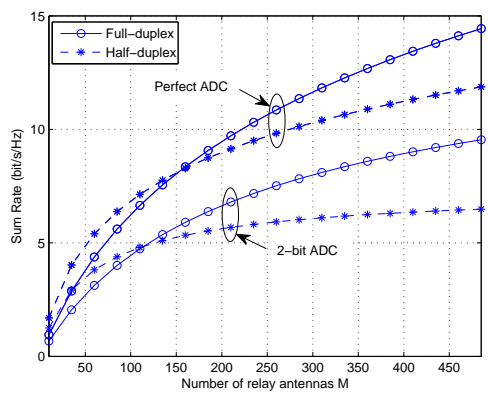}
    \caption{Sum rate versus $M$ for $K = 5$, $\sigma_\text{LI}^2 = 16$ dB, $p_\text p = 0$ dB, $p_\text S = 0$ dB, and $p_\text R = 0$ dB.}\label{fig:FD_vs_HD_num_antenna}
  \end{figure}

Fig. \ref{fig:FD_vs_HD_loop_inter} shows the sum rate versus the loop interference levels when the relay and destinations employ 2-bit ADCs. As expected, full-duplex outperforms half-duplex relaying at low $\sigma_\text{LI}^2$. However, when $\sigma_\text{LI}^2$ is high, loop interference dominates and hence the full-duplex mode is inferior. In such case, we can deploy more antennas to mitigate the effect of loop interference. For instance, by increasing the number of relay antennas from $M = 100$ to $M = 200$, the operating region where the full-duplex relaying works better enlarges, i.e., from $\sigma_\text{LI,0}^2 = 13.5$ dB to $\sigma_\text{LI,0}^2 = 15.5$ dB. This fact is further emphasized in Fig. \ref{fig:FD_vs_HD_num_antenna}. By focusing on the point $M = 185$ in Fig. \ref{fig:FD_vs_HD_num_antenna}, we observe that the half-duplex is superior with perfect ADCs while is inferior with 2-bit ADCs compared to the full-duplex mode, which indicates that the half-duplex is more sensitive to the low-resolution ADCs. The reason is that the quantization noise scales with the power of input signals, which becomes large since the transmit powers are doubled in the half-duplex mode.

\section{Conclusion}\label{sec:conclusion}
We analyzed the achievable sum rate of a multipair full-duplex massive antenna relaying system assuming that both the relay and the destinations use low-resolution ADCs. Exact and approximated closed-form expressions for the achievable sum rate were derived, based on which, the impact of key system parameters was characterized. The findings suggested that, deploying massive relay antenna arrays is an effective approach to compensate for the rate loss due to low-resolution ADCs at the relay, yet becomes ineffective to deal with the rate degradation due to the low-resolution ADCs at the destinations, which indicates that it is important to use higher resolution ADCs at the destination. In addition, it was revealed that, despite the use of low-resolution ADCs, employing massive antenna array at the relay enables significant power savings, i.e., the transmit power of each source can be scaled down proportional to $1/M$, to maintain a constant rate.
\appendices
\section{Proof of Theorem \ref{theorm:channel_esti}}\label{app:theorm:channel_esti}
We focus on the derivation of ${\bf{\hat G}}_\text{SR}$ only since ${\bf{\hat G}}_\text{RD}$ can be deduced in the same fashion.

To start with, we substitute \eqref{eq:Yrp} into \eqref{eq:Yrp:tilde}, and obtain
\begin{align}
  {\tilde{\bf Y}}_{\text{rp}}^H &= \alpha \sqrt{\tau_{\text p} p_{\text p}} {\bf \Phi}_{\text S}^H {\bf G}_{\text{SR}}^H + \alpha {\bf N}_{\text{rp}}^H + {{\bf N}}_{\text{rq}}^H.
\end{align}

Then, following the standard MMSE channel estimation technique, ${\bf{\hat G}}_\text{SR}^H$ is given by
\begin{align}\label{eq:G_SR:hat}
  {\bf{\hat G}}_\text{SR}^H = {\tt E} \left\{ {\bf{G}}_\text{SR}^H {\tilde{\bf Y}}_{\text{rp}} \right\} \left({\tt E}\left\{{\tilde{\bf Y}}_{\text{rp}}^H {\tilde{\bf Y}}_{\text{rp}} \right\} \right)^{-1} {\tilde{\bf Y}}_{\text{rp}}^H.
\end{align}

We first calculate
\begin{align}\label{eq:G_SR:hat:1}
  {\tt E} \left\{ {\bf{G}}_\text{SR}^H {\tilde{\bf Y}}_{\text{rp}} \right\}  &= {\tt E} \left\{ \alpha \sqrt{\tau_{\text p} p_{\text p}} {\bf{G}}_\text{SR}^H {\bf{G}}_\text{SR} {\bf \Phi}_{\text S} \right\} \\ \notag
  &= M \alpha \sqrt{\tau_{\text p} p_{\text p}} {\bf D}_\text{SR} {\bf \Phi}_{\text S}.
\end{align}
and
\begin{align}
    & {\tt E}\left\{{\tilde{\bf Y}}_{\text{rp}}^H {\tilde{\bf Y}}_{\text{rp}} \right\} \\ \notag
    &= {\tt E} \left\{ \alpha^2 \tau_{\text p} p_{\text p} {\bf \Phi}_{\text S}^H {\bf{G}}_\text{SR}^H {\bf{G}}_\text{SR} {\bf \Phi}_{\text S} \right\} + \alpha^2 {\bf I}_{\tau_\text p} + {\tt E} \left\{{{\bf N}}_{\text{rq}}^H {{\bf N}}_{\text{rq}} \right\}\\ \notag
    &= M \alpha^2 \tau_{\text p} p_{\text p} {\bf \Phi}_{\text S}^H {\bf{D}}_\text{SR}{\bf \Phi}_{\text S} + M \alpha^2 {\bf I}_{\tau_\text p} \\ \notag
    &+ \alpha \left(1 - \alpha\right)\text{diag} \left( {\tt E} \left\{{{\bf Y}}_{\text{rp}}^H {{\bf Y}}_{\text{rp}} \right\} \right)\\ \notag
    &= M \alpha^2 \tau_{\text p} p_{\text p} {\bf \Phi}_{\text S}^H {\bf{D}}_\text{SR}{\bf \Phi}_{\text S} + M \alpha^2 {\bf I}_{\tau_\text p} \\ \notag
    &+ M \alpha \left(1 - \alpha\right) \text{diag} \left(\tau_\text p p_\text p {\bf \Phi}_{\text S}^H {\bf{D}}_\text{SR}{\bf \Phi}_{\text S} + {\bf I}_{\tau_\text p} \right).
\end{align}

Here, we choose $\tau_\text p = K$, and assume that ${\bf \Phi}_{\text S}$ and ${\bf \Phi}_{\text D}$ are identity matrices, i.e., ${\bf \Phi}_{\text S} = {\bf \Phi}_{\text D} = {\bf I}_K$, for trackable analysis as in \cite{M.T.Ivrlac}. As a result, we have
\begin{align}
 \text{diag} \left({\bf \Phi}_{\text S}^H {\bf{D}}_\text{SR}{\bf \Phi}_{\text S}\right) = {\bf{D}}_\text{SR},
\end{align}
thus,
\begin{align}\label{eq:G_SR:hat:2}
{\tt E}\left\{{\tilde{\bf Y}}_{\text{rp}}^H {\tilde{\bf Y}}_{\text{rp}} \right\} = M \alpha \left(\tau_\text p p_\text p  {\bf{D}}_\text{SR} + {\bf I}_{\tau_\text p} \right).
\end{align}


Using the property of $\left({\bf A}^H {\bf A} + a {\bf I}_n\right)^{-1} {\bf A}^H = {\bf A}^H \left({\bf A} {\bf A}^H + a {\bf I}_m\right)^{-1}$ (${\bf A} \in {\mathbb C}^{m \times n} $), and then substituting \eqref{eq:G_SR:hat:1} and \eqref{eq:G_SR:hat:2} into \eqref{eq:G_SR:hat}, we have
\begin{align}
  {\bf{\hat G}}_\text{SR}^H = \alpha \left({\bf I}_K + \frac{{\bf D}_\text{SR}^{-1}}{\tau_\text p p_\text p} \right)^{-1} \left({\bf G}_\text{SR} + \frac{ {\bf N}_\text{rp}}{\sqrt{\tau_\text p p_\text p}} + \frac{ {\bf N}_\text{rq}}{\alpha\sqrt{\tau_\text p p_\text p}} \right).
\end{align}

As a result, we arrive the desired results after some simple mathematical derivations.
\section{Proof of Theorem \ref{theorm:se_exact}}\label{app:theorm:se_exact}
Capitalizing on the results of \cite{H.Q.Ngo2}, the achievable rate of $k$-th user is given by \eqref{eq:R_k}, where
$A_k =  p_{\text S} |{\tt E} \left\{{\bf g}_{{\text{RD}},k}^T {\bf F} {\bf g}_{{\text{SR}},k} \right\}|^2$ is the  desired signal power, $  B_k =  p_{\text S} \mathbb{V}{\text{ar}} \left({\bf g}_{{\text{RD}},k}^T {\bf F} {\bf g}_{{\text{SR}},k} \right)$ is the estimation error, $  C_k =  p_{\text S} \sum\limits_{j \neq k} {\tt E} \left\{|{\bf g}_{{\text{RD}},k}^T {\bf F} {\bf g}_{{\text{SR}},j}|^2 \right\}$ is the interpair interference, $D_k =  \frac{p_\text R}{M} {\tt E} \left\{ ||{\bf g}_{{\text{RD}},k}^T {\bf F} {\bf G}_{\text{RR}} ||^2 \right\}$, is the loop interference, $ E_k =  {\tt E} \left\{|| {\bf g}_{{\text{RD}},k}^T {\bf F} ||^2 \right\}$ is the noise at the relay, $  F_k = \frac{1}{\alpha^2} {\tt E} \left\{ |{\bf g}_{{\text{RD}},k}^T {\bf F} {\bf{\tilde n}}_{{\text R}}[i-d]|^2 \right\}$ is the quantization noise at the relay, $  G_k = \frac{1}{\alpha^2 \gamma^2}$ is the noise at the \emph{k}-th destination, $  H_k = \frac{1}{\alpha^2 \theta^2 \gamma^2}{\tt E} \left\{|{\tilde n}_{{\text D},k}[i]|^2 \right\}$ is the quantization noise at the \emph{k}-th destination. Besides these terms, we also need to calculate the normalization factor $\gamma$ in \eqref{eq:gamma}. In the following, we compute them one by one.

1) Compute $\gamma$:

(a) ${\tt E}\left\{||{\bf F} {\bf G}_\text{SR}||^2\right\} = {\tt E} \left\{{\text{tr}} \left({\bf G}_\text{SR}^H {\bf{\hat G}}_\text{SR} {\bf{\hat G}}_\text{RD}^T {\bf{\hat G}}_\text{RD}^* {\bf{\hat G}}_\text{SR}^H {\bf{G}}_\text{SR} \right) \right\} \\= {\text{tr}} \left( {\tt E} \left\{ {\bf{\hat G}}_\text{SR}^H {\bf{ G}}_\text{SR} {\bf{ G}}_\text{SR}^H {\bf{\hat G}}_\text{SR}  \right\} {\tt E}\left\{{\bf{\hat G}}_\text{RD}^T {\bf{\hat G}}_\text{RD}^* \right\}\right) =  M^3 \sum\limits_{i = 1}^K \sigma_{\text{SR},i}^4 \sigma_{\text{RD},i}^2 + M^2 \sum\limits_{i = 1}^K \sigma_{\text{SR},i}^2 \sigma_{\text{RD},i}^2 \sum\limits_{j = 1}^K \beta_{\text{SR},j}$.

(b) ${\tt E} \left\{||{\bf F} {\bf G}_\text{RR}||^2 \right\} = {\text{tr}} \left( {\tt E} \left\{{\bf G}_\text{RR} {\bf G}_\text{RR}^H \right\} {\tt E} \left\{ {\bf F}^H {\bf F} \right\} \right) = {\text{tr}} \left(M \sigma_\text{LI}^2 {\tt E} \left\{ {\bf F}^H {\bf F} \right\} \right) = M^3 \sigma_\text{LI}^2 \sum\limits_{i = 1}^K \sigma_{\text{SR},i}^2 \sigma_{\text{RD},i}^2$.

(c) ${\tt E} \left\{||{\bf F}||^2 \right\} = {\text{tr}} \left( {\tt E}\left\{{\bf{\hat G}}_\text{SR}^H {\bf{\hat G}}_\text{SR} \right\} {\tt E}\left\{{\bf{\hat G}}_\text{RD}^T {\bf{\hat G}}_\text{RD}^* \right\} \right) = M^2 \sum\limits_{i = 1}^K \sigma_{\text{SR},i}^2 \sigma_{\text{RD},i}^2$

(d) ${\tt E} \left\{||{\bf F} {\bf{\tilde n}}_\text{R} [i-d]||^2 \right\} = \alpha \left(1 - \alpha\right)  \\{\text{tr}} \left({\tt E} \left\{ {\bf{\hat G}}_\text{RD}^* {\bf{\hat G}}_\text{SR}^H {\text{diag}} \left( p_\text S {\bf{ G}}_\text{SR} {\bf{ G}}_\text{SR}^H + \frac{p_\text R {\bf{ G}}_\text{RR} {\bf{ G}}_\text{RR}^H }{M} + {\bf I}_M \right) {\bf{\hat G}}_\text{SR} {\bf{\hat G}}_\text{RD}^T  \right\}\right)$.

Firstly, we compute
\begin{align}
  &{\text{tr}} \left({\tt E} \left\{ {\bf{\hat G}}_\text{RD}^* {\bf{\hat G}}_\text{SR}^H {\text{diag}} \left({\bf{ G}}_\text{SR} {\bf{ G}}_\text{SR}^H \right) {\bf{\hat G}}_\text{SR} {\bf{\hat G}}_\text{RD}^T  \right\}\right) \\ \notag
  &= {\text{tr}} \left({\tt E} \left\{ {\bf{\hat G}}_\text{RD}^* {\bf{\hat G}}_\text{SR}^H {\text{diag}} \left({\bf{\hat G}}_\text{SR} {\bf{\hat G}}_\text{SR}^H \right) {\bf{\hat G}}_\text{SR} {\bf{\hat G}}_\text{RD}^T  \right\}\right) \\ \notag
  &+ {\text{tr}} \left({\tt E} \left\{ {\bf{\hat G}}_\text{RD}^* {\bf{\hat G}}_\text{SR}^H {\text{diag}} \left({\bf{ E}}_\text{SR} {\bf{ E}}_\text{SR}^H \right) {\bf{\hat G}}_\text{SR} {\bf{\hat G}}_\text{RD}^T  \right\}\right). \notag
\end{align}

By utilizing the fact that the channel matrices ${\bf{\hat G}}_\text{SR}$, ${\bf{ E}}_\text{SR}$, and ${\bf{\hat G}}_\text{RD}$ are independent of each other,
${\tt E} \left\{ {\bf{\hat G}}_\text{SR}^H {\text{diag}} \left({\bf{\hat G}}_\text{SR} {\bf{\hat G}}_\text{SR}^H \right) {\bf{\hat G}}_\text{SR} \right\} = M {\bf A}$ (where ${\bf A}$ is a $K \times K$ diagonal matrix with $\left[{\bf A}\right]_{kk} = \sigma_{\text{SR},k}^4 + \sigma_{\text{SR},k}^2 \sum\limits_{i=1}^K \sigma_{\text{SR},i}^2$), and ${\tt E} \left\{ {\text{diag}} \left( {\bf{ E}}_\text{SR} {\bf{ E}}_\text{SR}^H\right)\right\} = \sum\limits_{i=1}^K {\tilde\sigma}_{\text{SR},i}^2 {\bf I}_M $, we have
 \begin{align}\label{eq:Fn:1}
  & {\text{tr}} \left({\tt E} \left\{ {\bf{\hat G}}_\text{RD}^* {\bf{\hat G}}_\text{SR}^H {\text{diag}} \left({\bf{ G}}_\text{SR} {\bf{ G}}_\text{SR}^H \right) {\bf{\hat G}}_\text{SR} {\bf{\hat G}}_\text{RD}^T  \right\}\right) \\ \notag
  &= {\text{tr}} \left({\tt E} \left\{ {\bf{\hat G}}_\text{RD}^* {\tt E} \left\{ {\bf{\hat G}}_\text{SR}^H {\text{diag}} \left({\bf{\hat G}}_\text{SR} {\bf{\hat G}}_\text{SR}^H \right) {\bf{\hat G}}_\text{SR} \right\} {\bf{\hat G}}_\text{RD}^T  \right\}\right)\\ \notag
   &+ {\text{tr}} \left({\tt E} \left\{ {\bf{\hat G}}_\text{RD}^* {\tt E} \left\{ {\bf{\hat G}}_\text{SR}^H {\tt E} \left\{ {\text{diag}} \left({\bf{ E}}_\text{SR} {\bf{ E}}_\text{SR}^H \right) \right\} {\bf{\hat G}}_\text{SR} \right\} {\bf{\hat G}}_\text{RD}^T  \right\}\right)\\ \notag
  &= M^2 \left( \sum\limits_{j =1}^K \sigma_{\text{SR},j}^4 \sigma_{\text{RD},j}^2 + \sum\limits_{i =1}^K \sigma_{\text{SR},i}^2 \sum\limits_{j =1}^K \sigma_{\text{SR},j}^2 \sigma_{\text{RD},j}^2 \right) \\ \notag
  &+ M^2 \sum\limits_{i =1}^K {\tilde\sigma}_{\text{SR},i}^2 \sum\limits_{j =1}^K \sigma_{\text{SR},j}^2 \sigma_{\text{RD},j}^2 \\ \notag
  &= M^2 \left( \sum\limits_{j =1}^K \sigma_{\text{SR},j}^4 \sigma_{\text{RD},j}^2 + \sum\limits_{i =1}^K \beta_{\text{SR},i} \sum\limits_{j =1}^K \sigma_{\text{SR},j}^2 \sigma_{\text{RD},j}^2 \right).
 \end{align}

Then, following the same way for calculating ${\text{tr}} \left({\tt E} \left\{ {\bf{\hat G}}_\text{RD}^* {\bf{\hat G}}_\text{SR}^H {\text{diag}} \left({\bf{ G}}_\text{SR} {\bf{ G}}_\text{SR}^H \right) {\bf{\hat G}}_\text{SR} {\bf{\hat G}}_\text{RD}^T  \right\}\right)$, we obtain
\begin{align}\label{eq:Fn:2}
  &{\text{tr}} \left({\tt E} \left\{ {\bf{\hat G}}_\text{RD}^* {\bf{\hat G}}_\text{SR}^H {\text{diag}} \left({\bf{ G}}_\text{RR} {\bf{ G}}_\text{RR}^H \right) {\bf{\hat G}}_\text{SR} {\bf{\hat G}}_\text{RD}^T  \right\}\right) \\ \notag
  &=  M^3 \sigma_\text{LI}^2 \sum\limits_{j =1}^K \sigma_{\text{SR},j}^2 \sigma_{\text{RD},j}^2,\\
  &{\text{tr}} \left({\tt E} \left\{ {\bf{\hat G}}_\text{RD}^* {\bf{\hat G}}_\text{SR}^H {\bf{\hat G}}_\text{SR} {\bf{\hat G}}_\text{RD}^T  \right\}\right) =  M^2 \sum\limits_{j =1}^K \sigma_{\text{SR},j}^2 \sigma_{\text{RD},j}^2. \label{eq:Fn:3}
\end{align}

Substituting \eqref{eq:Fn:1}, \eqref{eq:Fn:2}, and \eqref{eq:Fn:3} into (d), we have
\begin{align}
  &{\tt E} \left\{||{\bf F} {\bf{\tilde n}}_\text{R}[i - d]||^2 \right\} \\ \notag
  &= \alpha \left(1 - \alpha \right) M^2 p_\text S \left( \sum\limits_{j =1}^K \sigma_{\text{SR},j}^4 \sigma_{\text{RD},j}^2 + \sum\limits_{i =1}^K \beta_{\text{SR},i} \sum\limits_{j =1}^K \sigma_{\text{SR},j}^2 \sigma_{\text{RD},j}^2 \right)\\
  &+ \alpha \left(1 - \alpha \right) M^2 \sum\limits_{j =1}^K \sigma_{\text{SR},j}^2 \sigma_{\text{RD},j}^2 \left(p_\text R \sigma_\text{LI}^2 + 1 \right).
\end{align}

To this end, combining the results (a), (b), (c), and (d), we obtain \eqref{eq:gamma:result}, shown on the top of the next page.
\begin{figure*}
\begin{align}\label{eq:gamma:result}
  \gamma = \frac{1}{M} \sqrt{\frac{p_\text R}{p_\text S \sum\limits_{i =1}^K \sigma_{\text{SR},i}^4 \sigma_{\text{RD},i}^2 \left(M \alpha^2 + \alpha \left(1 - \alpha\right) \right) + \alpha \sum\limits_{i =1}^K \sigma_{\text{SR},i}^2 \sigma_{\text{RD},i}^2 \left(p_\text S \sum\limits_{j =1}^K \beta_{\text{SR},j} + p_\text R \sigma_\text{LI}^2 + 1\right)}}.
\end{align}
\hrule
\end{figure*}

2) Calculate $A_k$:
\begin{align}
  &{\tt E} \left\{{\bf g}_{{\text{RD}},k}^T {\bf F} {\bf g}_{{\text{SR}},k} \right\} \\ \notag
  &= {\tt E} \left\{{\bf{\hat g}}_{{\text{RD}},k}^T {\bf{\hat g}}_{{\text{RD}},k}^* \right\} {\tt E} \left\{{\bf{\hat g}}_{{\text{SR}},k}^H {\bf{\hat g}}_{{\text{SR}},k} \right\} \\ \notag
  &= M^2 \sigma_{\text{SR},k}^2 \sigma_{\text{RD},k}^2.
\end{align}

Consequently, we have
\begin{align}\label{eq:Ak:result}
  { A}_k = \alpha^2 \theta^2 \gamma^2 p_{\text S} M^4 \sigma_{\text{SR},k}^4 \sigma_{\text{RD},k}^4.
\end{align}

3) Calculate $B_k$:
\begin{align}
  &{\tt E} \left\{ |{\bf g}_{{\text{RD}},k}^T {\bf F} {\bf g}_{{\text{SR}},k} |^2 \right\} \\ \notag
  &= {\tt E} \left\{\sum\limits_{n=1}^K \sum\limits_{l=1}^K {\bf g}_{{\text{RD}},k}^T {\bf{\hat g}}_{{\text{RD}},n}^* {\bf{\hat g}}_{{\text{SR}},n}^H {\bf g}_{{\text{SR}},k} {\bf g}_{{\text{SR}},k}^H {\bf{\hat g}}_{{\text{SR}},l} {\bf{\hat g}}_{{\text{RD}},l}^T {\bf g}_{{\text{RD}},k}^* \right\},
\end{align}
which can be decomposed into three different cases:

a) for $n \neq l \neq k$, we have ${\tt E} \left\{ |{\bf g}_{{\text{RD}},k}^T {\bf F} {\bf g}_{{\text{SR}},k} |^2 \right\} = 0$.

b) for $n = l \neq k$, we have
\begin{align}
  &{\tt E} \left\{ |{\bf g}_{{\text{RD}},k}^T {\bf F} {\bf g}_{{\text{SR}},k} |^2 \right\} \\ \notag
  &= {\tt E} \left\{\sum\limits_{n=1}^K {\bf g}_{{\text{RD}},k}^T {\bf{\hat g}}_{{\text{RD}},n}^* {\bf{\hat g}}_{{\text{SR}},n}^H {\bf g}_{{\text{SR}},k} {\bf g}_{{\text{SR}},k}^H {\bf{\hat g}}_{{\text{SR}},n} {\bf{\hat g}}_{{\text{RD}},n}^T {\bf g}_{{\text{RD}},k}^* \right\}\\  \notag
  &= M^2 \beta_{\text{SR},k} \beta_{\text{RD},k} \sum\limits_{n \neq k} \sigma_{\text{SR},n}^2 \sigma_{\text{RD},n}^2.
\end{align}

c) for $n = l = k$, we have
\begin{align}
  &{\tt E} \left\{ |{\bf g}_{{\text{RD}},k}^T {\bf F} {\bf g}_{{\text{SR}},k} |^2 \right\} \\ \notag
   &= {\tt E} \left\{{\bf g}_{{\text{RD}},k}^T {\bf{\hat g}}_{{\text{RD}},k}^* {\bf{\hat g}}_{{\text{SR}},k}^H {\bf g}_{{\text{SR}},k} {\bf g}_{{\text{SR}},k}^H {\bf{\hat g}}_{{\text{SR}},k} {\bf{\hat g}}_{{\text{RD}},k}^T {\bf g}_{{\text{RD}},k}^* \right\} \\ \notag
  &= {\tt E} \left\{{\bf{\hat g}}_{{\text{RD}},k}^T {\bf{\hat g}}_{{\text{RD}},k}^* {\bf{\hat g}}_{{\text{SR}},k}^H {\bf{\hat g}}_{{\text{SR}},k} {\bf{\hat g}}_{{\text{SR}},k}^H {\bf{\hat g}}_{{\text{SR}},k} {\bf{\hat g}}_{{\text{RD}},k}^T {\bf{\hat g}}_{{\text{RD}},k}^* \right\} \\
  &+ {\tt E} \left\{{\bf{\hat g}}_{{\text{RD}},k}^T {\bf{\hat g}}_{{\text{RD}},k}^* {\bf{\hat g}}_{{\text{SR}},k}^H {\bf e}_{{\text{SR}},k} {\bf e}_{{\text{SR}},k}^H {\bf{\hat g}}_{{\text{SR}},k} {\bf{\hat g}}_{{\text{RD}},k}^T {\bf{\hat g}}_{{\text{RD}},k}^* \right\}\\ \notag
  &+ {\tt E} \left\{{\bf e}_{{\text{RD}},k}^T {\bf{\hat g}}_{{\text{RD}},k}^* {\bf{\hat g}}_{{\text{SR}},k}^H {\bf{\hat g}}_{{\text{SR}},k} {\bf{\hat g}}_{{\text{SR}},k}^H {\bf{\hat g}}_{{\text{SR}},k} {\bf{\hat g}}_{{\text{RD}},k}^T {\bf e}_{{\text{RD}},k}^* \right\} \\
  &+ {\tt E} \left\{{\bf e}_{{\text{RD}},k}^T {\bf{\hat g}}_{{\text{RD}},k}^* {\bf{\hat g}}_{{\text{SR}},k}^H {\bf e}_{{\text{SR}},k} {\bf e}_{{\text{SR}},k}^H {\bf{\hat g}}_{{\text{SR}},k} {\bf{\hat g}}_{{\text{RD}},k}^T {\bf e}_{{\text{RD}},k}^* \right\}\\ \notag
  &= M^2 \left(M + 1 \right)^2 \sigma_{\text{SR},k}^4 \sigma_{\text{RD},k}^4 + M^2 \left(M + 1 \right) \sigma_{\text{SR},k}^2 {\tilde\sigma}_{\text{SR},k}^2 \sigma_{\text{RD},k}^4 \\ \notag
  &+ M^2 \left(M + 1 \right) \sigma_{\text{SR},k}^4 \sigma_{\text{RD},k}^2 {\tilde\sigma}_{\text{RD},k}^2 + M^2 \sigma_{\text{SR},k}^2 {\tilde\sigma}_{\text{SR},k}^2 \sigma_{\text{RD},k}^2 {\tilde\sigma}_{\text{RD},k}^2 .
\end{align}

Finally, combining a), b), and c), we obtain
\begin{align}\label{eq:Bk:result}
  &{B}_k = \alpha^2 \theta^2 \gamma^2 p_{\text S} \left( {\tt E} \left\{ |{\bf g}_{{\text{RD}},k}^T {\bf F} {\bf g}_{{\text{SR}},k} |^2 \right\} - |{\tt E} \left\{{\bf g}_{{\text{RD}},k}^T {\bf F} {\bf g}_{{\text{SR}},k} \right\}|^2 \right) \\ \notag
  &= \alpha^2 \theta^2 \gamma^2 p_{\text S} M^3 \sigma_{\text{SR},k}^2 \sigma_{\text{RD},k}^2 \left(\beta_{\text{SR},k} \sigma_{\text{RD},k}^2 + \beta_{\text{RD},k} \sigma_{\text{SR},k}^2 \right) \\ \notag
  &\alpha^2 \theta^2 \gamma^2 p_{\text S} M^2 \beta_{\text{SR},k} \beta_{\text{RD},k} \sum\limits_{n =1}^K \sigma_{\text{SR},n}^2 \sigma_{\text{RD},n}^2.
\end{align}

4) Calculate $C_k$:

We first rewrite ${\tt E} \left\{ |{\bf g}_{{\text{RD}},k}^T {\bf F} {\bf g}_{{\text{SR}},j} |^2 \right\}$ for $j \neq k$ as
\begin{align}\label{eq:Ck:prime}
  &{\tt E} \left\{ |{\bf g}_{{\text{RD}},k}^T {\bf F} {\bf g}_{{\text{SR}},j} |^2 \right\} \\ \notag
   &= {\tt E} \left\{\sum\limits_{n=1}^K \sum\limits_{l=1}^K {\bf g}_{{\text{RD}},k}^T {\bf{\hat g}}_{{\text{RD}},n}^* {\bf{\hat g}}_{{\text{SR}},n}^H {\bf g}_{{\text{SR}},j} {\bf g}_{{\text{SR}},j}^H {\bf{\hat g}}_{{\text{SR}},l} {\bf{\hat g}}_{{\text{RD}},l}^T {\bf g}_{{\text{RD}},k}^* \right\}.
\end{align}

Next, \eqref{eq:Ck:prime} can be split into six different cases:

a) for $n \neq l \neq k$, we have ${\tt E} \left\{ |{\bf g}_{{\text{RD}},k}^T {\bf F} {\bf g}_{{\text{SR}},j} |^2 \right\} = 0$.

b) for $n = l \neq k$, we have ${\tt E} \left\{ |{\bf g}_{{\text{RD}},k}^T {\bf F} {\bf g}_{{\text{SR}},j} |^2 \right\} =  M^2 \beta_{\text{SR},j} \beta_{\text{RD},k} \sum\limits_{n\neq k,j}^K \sigma_{\text{SR},n}^2 \sigma_{\text{RD},n}^2$.

c) for $n = l = k$, we have ${\tt E} \left\{ |{\bf g}_{{\text{RD}},k}^T {\bf F} {\bf g}_{{\text{SR}},j} |^2 \right\} =  M^2 \beta_{\text{SR},j} \sigma_{\text{SR},k}^2 \sigma_{\text{RD},k}^2 \left(M \sigma_{\text{RD},k}^2 + \beta_{\text{RD},k} \right) $.

d) for $n = l = j$, we have ${\tt E} \left\{ |{\bf g}_{{\text{RD}},k}^T {\bf F} {\bf g}_{{\text{SR}},j} |^2 \right\} =  M^2 \beta_{\text{RD},k} \sigma_{\text{SR},j}^2 \sigma_{\text{RD},j}^2 \left(M \sigma_{\text{SR},j}^2 + \beta_{\text{SR},j} \right) $.

e) for $n =k, l = j$, we have ${\tt E} \left\{ |{\bf g}_{{\text{RD}},k}^T {\bf F} {\bf g}_{{\text{SR}},j} |^2 \right\} = 0$.

f) for $n =j, l = k$, we have ${\tt E} \left\{ |{\bf g}_{{\text{RD}},k}^T {\bf F} {\bf g}_{{\text{SR}},j} |^2 \right\} = 0$.

Altogether, $C_k$ is given by
\begin{align}\label{eq:Ck:result}
  &C_k = \alpha^2 \theta^2 \gamma^2 p_{\text S} M^3 \sum\limits_{j \neq k} \left( \sigma_{\text{SR},k}^2 \sigma_{\text{RD},k}^4 \beta_{\text{SR},j} + \beta_{\text{RD},k} \sigma_{\text{SR},j}^4 \sigma_{\text{RD},j}^2  \right) \\ \notag
  & + \alpha^2 \theta^2 \gamma^2 p_{\text S} M^2 \sum\limits_{j \neq k} \beta_{\text{SR},j} \beta_{\text{RD},k} \sum\limits_{n\neq k,j}^K \sigma_{\text{SR},n}^2 \sigma_{\text{RD},n}^2 \\ \notag
  &+ \alpha^2 \theta^2 \gamma^2 p_{\text S} M^2 \sum\limits_{j \neq k} \beta_{\text{SR},j} \beta_{\text{RD},k} \sigma_{\text{SR},k}^2 \sigma_{\text{RD},k}^2 \\ \notag
  &+ \alpha^2 \theta^2 \gamma^2 p_{\text S} M^2 \sum\limits_{j \neq k}  + \beta_{\text{SR},j} \beta_{\text{RD},k} \sigma_{\text{SR},j}^2 \sigma_{\text{RD},j}^2.
\end{align}

5) Calculate $D_k$ and $E_k$:

Following the same technique for deriving $B_k$, we can obtain
\begin{align}\label{eq:Hk:2}
&{\tt E} \left\{ ||{\bf g}_{{\text{RD}},k}^T {\bf F} {\bf G}_{\text{RR}} ||^2 \right\} \\ \notag
 &= M^3 \sigma_\text{LI}^2 \left( M \sigma_{\text{SR},k}^2 \sigma_{\text{RD},k}^4 + \beta_{\text{RD},k} \sum\limits_{n=1}^K \sigma_{\text{SR},n}^2 \sigma_{\text{RD},n}^2 \right),\\ \notag
&{\tt E} \left\{|| {\bf g}_{{\text{RD}},k}^T {\bf F} ||^2 \right\} \\
&= M^2 \left( M \sigma_{\text{SR},k}^2 \sigma_{\text{RD},k}^4 + \beta_{\text{RD},k} \sum\limits_{n=1}^K \sigma_{\text{SR},n}^2 \sigma_{\text{RD},n}^2 \right). \label{eq:Hk:3}
\end{align}

Thus, $D_k$ and $E_k$ are given by
\begin{align}\label{eq:Dk:result}
  &D_k = \alpha^2 \theta^2 \gamma^2 M^3 \sigma_\text{LI}^2 p_\text R \sigma_{\text{SR},k}^2 \sigma_{\text{RD},k}^4 \\ \notag
  &+\alpha^2 \theta^2 \gamma^2 M^2 \sigma_\text{LI}^2 p_\text R \beta_{\text{RD},k} \sum\limits_{n=1}^K \sigma_{\text{SR},n}^2 \sigma_{\text{RD},n}^2,\\ \notag
  &E_k= \\
  &\alpha^2 \theta^2 \gamma^2 M^2 \left( M \sigma_{\text{SR},k}^2 \sigma_{\text{RD},k}^4 + \beta_{\text{RD},k} \sum\limits_{n=1}^K \sigma_{\text{SR},n}^2 \sigma_{\text{RD},n}^2 \right).\label{eq:Ek:result}
\end{align}

6) Calculate $F_k$:

By using the fact that the channel matrices ${\bf G}_\text{RD}$ and ${\bf {\hat G}}_\text{RD}$ are independent of ${\bf G}_\text{SR}$, ${\bf{\hat G}}_\text{SR}$, and ${\bf G}_\text{RR}$, we have
\begin{align}\label{eq:Fk}
  &{\tt E} \left\{ |{\bf g}_{{\text{RD}},k}^T {\bf F} {\bf{\tilde n}}_{{\text R}}[i-d]|^2 \right\}\\ \notag
  &= \alpha \left(1 - \alpha\right) {\tt E} \left\{{\bf g}_{\text{RD},k}^T {\bf {\hat G}}_\text{RD}^* {\tt E}\left\{{\bf B}\right\} {\bf {\hat G}}_\text{RD}^T {\bf g}_{\text{RD},k}^* \right\},
\end{align}
where
\begin{align}
&{\tt E}\left\{{\bf B}\right\} = \\ \notag
&{\tt E} \left\{ {\bf{\hat G}}_\text{SR}^H {\text{diag}} \left( p_\text S {\bf{ G}}_\text{SR} {\bf{ G}}_\text{SR}^H + \frac{p_\text R {\bf{ G}}_\text{RR} {\bf{ G}}_\text{RR}^H }{M} + {\bf I}_M \right) {\bf{\hat G}}_\text{SR}\right\}.
\end{align}

Then, following the same fashion as for deducing ${\tt E} \left\{||{\bf F} {\bf{\tilde n}}_\text{R} [i-d]||^2 \right\}$, we can easily obtain
\begin{align}\label{eq:Fk:1}
    &{\tt E} \left\{{\bf g}_{\text{RD},k}^T {\bf {\hat G}}_\text{RD}^* {\bf{\hat G}}_\text{SR}^H {\text{diag}} \left({\bf{ G}}_\text{SR} {\bf{ G}}_\text{SR}^H \right) {\bf{\hat G}}_\text{SR} {\bf {\hat G}}_\text{RD}^T {\bf g}_{\text{RD},k}^* \right\} \\ \notag
     &= M^2 \beta_{\text{RD},k} \sum\limits_{n\neq k} \sigma_{\text{SR},n}^2 \sigma_{\text{RD},n}^2 \left(\sigma_{\text{SR},n}^2 + \sum\limits_{i=1}^K\beta_{\text{SR},i}\right) \\ \notag
    &+M^2 \sigma_{\text{SR},k}^2 \sigma_{\text{RD},k}^2 \left(M \sigma_{\text{RD},k}^2 + \beta_{\text{RD},k} \right)\left(\sigma_{\text{SR},k}^2 + \sum\limits_{i=1}^K\beta_{\text{SR},i}\right), \\ \label{eq:Fk:2}
    &{\tt E} \left\{{\bf g}_{\text{RD},k}^T {\bf {\hat G}}_\text{RD}^* {\bf{\hat G}}_\text{SR}^H {\text{diag}} \left({\bf{ G}}_\text{RR} {\bf{ G}}_\text{RR}^H \right) {\bf{\hat G}}_\text{SR} {\bf {\hat G}}_\text{RD}^T {\bf g}_{\text{RD},k}^* \right\} \\ \notag
    &= M^3 \sigma_\text{LI}^2 \left(M \sigma_{\text{SR},k}^2 \sigma_{\text{RD},k}^4 + \beta_{\text{RD},k} \sum\limits_{n=1}^K\sigma_{\text{SR},n}^2 \sigma_{\text{RD},n}^2 \right), \\ \label{eq:Fk:3}
    &{\tt E} \left\{{\bf g}_{\text{RD},k}^T {\bf {\hat G}}_\text{RD}^* {\bf{\hat G}}_\text{SR}^H {\bf{\hat G}}_\text{SR} {\bf {\hat G}}_\text{RD}^T {\bf g}_{\text{RD},k}^* \right\} \\ \notag
    &= M^3 \sigma_{\text{SR},k}^2 \sigma_{\text{RD},k}^4 + M^2 \beta_{\text{RD},k} \sum\limits_{n=1}^K\sigma_{\text{SR},n}^2 \sigma_{\text{RD},n}^2.
\end{align}

Substituting \eqref{eq:Fk:1}, \eqref{eq:Fk:2}, and \eqref{eq:Fk:3} into \eqref{eq:Fk}, we have
\begin{align}\label{eq:Fk:result}
  F_k &= \frac{1-\alpha}{\alpha} M^3  p_{\text S} \sigma_{\text{SR},k}^2 \sigma_{\text{RD},k}^4 \left( \sigma_{\text{SR},k}^2 + \sum\limits_{i =1}^K \beta_{\text{SR},i} \right)\\ \notag
   & + \frac{1-\alpha}{\alpha} M^3 \left(p_\text R \sigma_\text{LI}^2 + 1\right) \sigma_{\text{SR},k}^2 \sigma_{\text{RD},k}^4\\ \notag
   &+ \frac{1-\alpha}{\alpha} M^2 \left(p_\text R \sigma_\text{LI}^2 + 1\right) \beta_{\text{RD},k} \sum\limits_{n=1}^K\sigma_{\text{SR},n}^2 \sigma_{\text{RD},n}^2 \\ \notag
   &+ \frac{1-\alpha}{\alpha} M^2 p_\text S \beta_{\text{RD},k} \sum\limits_{n=1}^K \sigma_{\text{SR},n}^2 \sigma_{\text{RD},n}^2 \left(\sigma_{\text{SR},n}^2 + \sum\limits_{i=1}^K\beta_{\text{SR},i}\right).
\end{align}

7) Calculate $H_k$:
\begin{align}\label{eq:Hk}
  &{\tt E} \left\{|{\tilde n}_{{\text D},k}[i]|^2 \right\} = \theta \left(1 - \theta \right) {\tt E} \left\{|{ y}_{{\text D},k}[i]|^2 \right\} \\ \notag
  &= \alpha^2 \theta \left(1 - \theta\right)\gamma^2 p_\text S {\tt E}\left\{||{\bf g}_{\text{RD},k}^T{\bf F}{\bf G}_\text{SR}||^2 \right\} \\ \notag
   &+ \alpha^2 \theta \left(1 - \theta\right)\gamma^2 \frac{p_\text R}{M}  {\tt E}\left\{||{\bf g}_{\text{RD},k}^T{\bf F}{\bf G}_\text{RR}||^2 \right\} \\ \notag
  &+ \alpha^2 \theta \left(1 - \theta\right)\gamma^2  {\tt E}\left\{||{\bf g}_{\text{RD},k}^T{\bf F}||^2 \right\} \\ \notag
  &+ \theta \left(1 - \theta\right)\gamma^2  {\tt E}\left\{|{\bf g}_{\text{RD},k}^T{\bf F} {\bf{\tilde n}}_{{\text R}}[i-d]|^2 \right\} + \theta \left(1 - \theta\right).
\end{align}

We first calculate
\begin{align}\label{eq:Hk:1}
 &{\tt E}\left\{||{\bf g}_{\text{RD},k}^T{\bf F}{\bf G}_\text{SR}||^2 \right\}\\ \notag
 &= M^3\sigma_{\text{SR},k}^2 \sigma_{\text{RD},k}^4 \left(M\sigma_{\text{SR},k}^2 + \sum\limits_{i=1}^K \beta_{\text{SR},i}\right)  \\ \notag
 &M^2\sigma_{\text{SR},k}^2 \left(M\sigma_{\text{SR},k}^2 + \sum\limits_{i=1}^K \beta_{\text{SR},i}\right) \beta_{\text{RD},k} \sum\limits_{i=1}^K \sigma_{\text{RD},i}^2 .
\end{align}

Then, by substituting \eqref{eq:Hk:1}, \eqref{eq:Hk:2}, \eqref{eq:Hk:3}, and \eqref{eq:Fk} into \eqref{eq:Hk}, we arrive the following result:
\begin{align}\label{eq:Hk:result}
      H_k &= \frac{1-\theta}{\theta} p_{\text S} M^3\sigma_{\text{SR},k}^2 \sigma_{\text{RD},k}^4 \left(M\sigma_{\text{SR},k}^2 + \sum\limits_{i=1}^K \beta_{\text{SR},i}\right) \\ \notag
  &+ \frac{1-\theta}{\theta} p_{\text S} M^2\sigma_{\text{SR},k}^2 \left(M\sigma_{\text{SR},k}^2 + \sum\limits_{i=1}^K \beta_{\text{SR},i}\right)\beta_{\text{RD},k} \sum\limits_{i=1}^K \sigma_{\text{RD},i}^2 \\ \notag
  &+  \frac{1-\theta}{\alpha \theta} M^3 \left(p_\text R \sigma_\text{LI}^2 + 1\right) \sigma_{\text{SR},k}^2 \sigma_{\text{RD},k}^4\\ \notag
  &+ \frac{1-\theta}{\alpha \theta} M^2 \left(p_\text R \sigma_\text{LI}^2 + 1\right) \beta_{\text{RD},k} \sum\limits_{n=1}^K \sigma_{\text{SR},n}^2 \sigma_{\text{RD},n}^2 \\ \notag
    &+  \frac{\left(1 - \alpha\right)\left(1-\theta\right)}{\alpha \theta} M^3  p_{\text S} \sigma_{\text{SR},k}^2 \sigma_{\text{RD},k}^4 \left( \sigma_{\text{SR},k}^2 + \sum\limits_{i =1}^K \beta_{\text{SR},i} \right)\\ \notag
   &+  \frac{\left(1 - \alpha\right)\left(1-\theta\right)}{\alpha \theta}  p_\text S M^2 \beta_{\text{RD},k} \sum\limits_{n=1}^K \sigma_{\text{SR},n}^4 \sigma_{\text{RD},n}^2 \\ \notag
   &+ \frac{\left(1 - \alpha\right)\left(1-\theta\right)}{\alpha \theta}  p_\text S M^2 \beta_{\text{RD},k} \sum\limits_{n=1}^K \sigma_{\text{SR},n}^2 \sigma_{\text{RD},n}^2 \sum\limits_{i=1}^K\beta_{\text{SR},i} \\ \notag
   &+ \frac{1-\theta}{\alpha^2 \theta \gamma^2}.
  \end{align}
Finally, combining \eqref{eq:gamma:result}, \eqref{eq:Ak:result}, \eqref{eq:Bk:result}, \eqref{eq:Ck:result}, \eqref{eq:Dk:result}, \eqref{eq:Ek:result}, \eqref{eq:Fk:result}, \eqref{eq:Hk:result}, and \eqref{eq:R_k}, and after some simple algebraic manipulation, we complete the proof.
\bibliographystyle{IEEE}
\begin{footnotesize}

 \end{footnotesize}

\end{document}